\documentclass[12pt,a4]{article}

\usepackage{amsthm,amsmath,latexsym,multibox,amssymb,amsfonts}
\usepackage{graphicx,lscape,fancyhdr,array,stmaryrd}

\newcommand{\bep}{\begin{picture}}
\newcommand{\eep}{\end{picture}}

\newcommand{\ArbitraryYoungDiagram}{{\bep(10,10)%
\unitlength=0.45mm\put(0,-77){\BlockCEnum{6}{3}{4}{2}{1}{2}{s}}%
\reptInit{1}\multiput(0,-2)(10,0){6}{\reptbyone{h}}%
\eep}}

\newcommand{\GaugeParameters}{{\bep(160,150){%
\put(0,0){\RectT{5}{4}{\TextTopLeftIn{\ensuremath{s_N-1}}{\ensuremath{p_N}}}}%
\put(50,20){\RectT{1}{2}{\TextRight{$k_N$}}}%
\put(50,0){\BlockADotted{1}{2}}%
\put(0,40){\RectT{9}{5}{\TextTopLeftIn{\ensuremath{s_2-1}}{\ensuremath{p_2}}}}%
\put(90,60){\RectT{1}{3}{\TextRight{$k_2$}}}%
\put(90,40){\BlockADotted{1}{2}}%
\put(0,90){\RectT{14}{6}{\TextTopLeftIn{\ensuremath{s_1-1}}{\ensuremath{p_1}}}}%
\put(140,120){\RectT{1}{3}{\TextRight{$k_1$}}}%
\put(140,90){\BlockADotted{1}{3}}%
}\eep}}

\newcounter{YoungHeight}\newcounter{YoungWidth}

\newcounter{Mul1}\newcounter{Mul2}\newcounter{Mul3}\newcounter{Mul4}
\newcounter{A0}\newcounter{A1}\newcounter{A2}\newcounter{A3}\newcounter{A4}\newcounter{A5}\newcounter{A6}
\newcounter{B0}\newcounter{B1}\newcounter{B2}\newcounter{B3}
\newcounter{C1}\newcounter{C2}\newcounter{C3}\newcounter{C4}\newcounter{C6}\newcounter{C7}
\newcounter{D1}\newcounter{D2}\newcounter{D3}\newcounter{D4}\newcounter{D5}
\newcounter{T0}\newcounter{T1}

\newcounter{TGR0}
\newcounter{R0}\newcounter{R1}\newcounter{R2}\newcounter{R3}
\newcounter{AR0}\newcounter{AR1}\newcounter{AR2}\newcounter{AR3}\newcounter{AR5}
\newcounter{Dotted0}\newcounter{Dotted1}\newcounter{Dotted2}\newcounter{Dotted3}

\newcounter{reptA}

\newlength{\txtHShift}

\newlength{\txtWidth}

\newcommand{\HalfLength}[2]{\setcounter{Mul1}{#1}\setcounter{Mul2}{#1}\addtocounter{Mul1}{\value{Mul2}}\addtocounter{Mul1}{\value{Mul2}}%
\addtocounter{Mul1}{\value{Mul2}}\addtocounter{Mul1}{\value{Mul2}}\setcounter{#2}{\value{Mul1}}}

\newcommand{\Add}[3]{\setcounter{#1}{#2}\addtocounter{#1}{#3}}

\newcommand{\Length}[1]{#10}
\newcommand{\YoungScale}{}

\newcommand{\reptInit}[1]{{\setcounter{reptA}{#1}}}

\newcommand{\reptbyone}[1]{{\ensuremath{{#1}_{\arabic{reptA}}}\addtocounter{reptA}{1}}}

\newcommand{\shiftedText}[2]{{\hspace{#1}#2}}
\newcommand{\calcHShift}[1]{\settowidth{\txtWidth}{#1}\setlength{\txtHShift}{-0.5\txtWidth}}

\newcommand{\TextTop}[3]{{\calcHShift{#1}\HalfLength{#2}{T0}\Add{T1}{\Length{#3}}{-9}\put(\value{T0},\value{T1}){\shiftedText{\txtHShift}{#1}}}}

\newcommand{\TextLeftIn}[3]{{\HalfLength{#3}{T1}\addtocounter{T1}{-5}\put(2,\value{T1}){#1}}}

\newcommand{\TextRight}[3]{{\HalfLength{#3}{T1}\addtocounter{T1}{-5}\Add{T0}{\Length{#2}}{2}\put(\value{T0},\value{T1}){#1}}}

\newcommand{\TextTopLeftIn}[4]{{\TextTop{#1}{#3}{#4}\TextLeftIn{#2}{#3}{#4}}}

\newcommand{\dottedline}[2]{{\HalfLength{#1}{Dotted0}\HalfLength{#2}{Dotted1}\Add{Dotted2}{#1}{#2}\HalfLength{\value{Dotted2}}{Dotted3}%
\qbezier[\value{Dotted3}](0,0)(\value{Dotted0},\value{Dotted1})(\Length{#1},\Length{#2})}}

\newcommand{\BlockA}[2]{{\YoungScale\bep(\Length{#1},\Length{#2}){\Add{A1}{#1}{1}\Add{A2}{#2}{1}}%
\multiput(0,0)(10,0){\value{A1}}{\line(0,1){\Length{#2}}}\multiput(0,0)(0,10){\value{A2}}{\line(1,0){\Length{#1}}}%
\setcounter{YoungHeight}{\Length{#2}}\setcounter{YoungWidth}{\Length{#1}}\eep}}

\newcommand{\BlockB}[4]{{\YoungScale\Add{B3}{\Length{#2}}{\Length{#4}}%
\bep(\Length{#1},\value{B3})\put(0,\Length{#4}){\BlockA{#1}{#2}}%
\put(0,0){\BlockA{#3}{#4}}\setcounter{YoungHeight}{\value{B3}}\setcounter{YoungWidth}{\Length{#1}}\eep}}

\newcommand{\BlockADotted}[2]{{\YoungScale\bep(\Length{#1},\Length{#2}){\Add{A1}{#1}{1}\Add{A2}{#2}{1}}%
\multiput(0,0)(10,0){\value{A1}}{\dottedline{0}{#2}}\multiput(0,0)(0,10){\value{A2}}{\dottedline{#1}{0}}%
\setcounter{YoungHeight}{\Length{#2}}\setcounter{YoungWidth}{\Length{#1}}\eep}}

\newcommand{\RectT}[3]{\bep(\Length{#1},\Length{#2})\put(0,0){\line(1,0){\Length{#1}}}\put(0,0){\line(0,1){\Length{#2}}}%
\put(\Length{#1},\Length{#2}){\line(-1,0){\Length{#1}}}\put(\Length{#1},\Length{#2}){\line(0,-1){\Length{#2}}}#3{#1}{#2}\eep}

\newcommand{\RectARow}[2]{{\bep(\Length{#1},10)\put(0,0){\RectT{#1}{1}{\TextTop{#2}}}\eep}}

\newcommand{\RectBRow}[4]{{\bep(\Length{#1},20)\put(0,0){\RectT{#2}{1}{\TextTop{#4}}}%
\put(0,10){\RectT{#1}{1}{\TextTop{#3}}}\eep}}

\newcommand{\RectAYoung}[3]{{\bep(0,0)#3\eep\Add{A0}{\value{YoungWidth}}{\Length{#1}}%
\Add{A1}{\value{YoungHeight}}{-10}\bep(\value{A0},\value{YoungHeight})%
\put(\value{YoungWidth},\value{A1}){\RectT{#1}{1}{\TextTop{#2}}}\eep}}

\newcommand{\RectBYoung}[3]{{\bep(0,0)\put(0,0){#3}\eep\Add{A0}{\value{YoungHeight}}{10}%
\bep(\Length{#1},\value{A0})%
\put(0,\value{YoungHeight}){\RectT{#1}{1}{\TextTop{#2}}}\eep}}

\newcommand{\RectCYoung}[5]{{\bep(0,0)\put(0,0){#5}\eep\Add{A0}{\value{YoungHeight}}{20}%
\bep(\Length{#1},\value{A0})%
\put(0,\value{YoungHeight}){\RectBRow{#1}{#2}{#3}{#4}}\eep}}

\newcommand{\BlockAEnum}[4]{{\YoungScale\Add{A1}{#1}{1}\Add{A2}{#2}{1}\Add{A3}{\Length{#1}}{5}\Add{A4}{\Length{#2}}{-10}\bep(\value{A3},\Length{#2}){%
\multiput(0,0)(10,0){\value{A1}}{\line(0,1){\Length{#2}}}\multiput(0,0)(0,10){\value{A2}}{\line(1,0){\Length{#1}}}%
\reptInit{#3}\multiput(\value{A3},\value{A4})(0,-10){#2}{\reptbyone{#4}}
}\eep}}

\newcommand{\BlockBEnum}[5]{{\YoungScale\Add{B2}{\Length{#1}}{10}\Add{B1}{#2}{1}\Add{B3}{\Length{#2}}{\Length{#4}}\bep(\value{B2},\value{B3})\put(0,\Length{#4}){\BlockAEnum{#1}{#2}{1}{#5}}%
\put(0,0){\BlockAEnum{#3}{#4}{\value{B1}}{#5}}\eep}}

\newcommand{\BlockCEnum}[7]{{\YoungScale\Add{C2}{\Length{#1}}{10}\Add{C1}{#2}{#4}\addtocounter{C1}{1}%
\Add{C3}{\Length{#4}}{\Length{#6}}\Add{C4}{\value{C3}}{\Length{#2}}\bep(\value{C2},\value{C4})%
\put(0,\Length{#6}){\BlockBEnum{#1}{#2}{#3}{#4}{#7}}\put(0,0){\BlockAEnum{#5}{#6}{\value{C1}}{#7}}\eep}}

\newcommand{\YoungA}{\BlockA{1}{1}}
\newcommand{\YoungB}{\BlockA{2}{1}}

\newcommand{\YoungAA}{\BlockA{1}{2}}
\newcommand{\YoungBA}{\BlockB{2}{1}{1}{1}}
\newcommand{\YoungBB}{\BlockA{2}{2}}

\usepackage{amsthm,amsmath,latexsym,multibox,amssymb,amsfonts}
\usepackage{graphicx,lscape,fancyhdr,array,stmaryrd,euscript,wrapfig}

\renewcommand{\theequation}{\arabic{section}.\arabic{equation}}

\newcommand{\AlgebraFont}[1]{\mathfrak{#1}}

\newcommand{\lorentz}{{\ensuremath{\AlgebraFont{so}(d-1,1)}}}
\newcommand{\msv}{{\ensuremath{\AlgebraFont{so}(d-1)}}}
\newcommand{\mls}{{\ensuremath{\AlgebraFont{so}(d-2)}}}

\newcommand{\isoLittle}{{\ensuremath{\AlgebraFont{iso}(d-2)}}}

\newcommand{\pl}{\partial}

\newcommand{\be}{\begin{equation}}
\newcommand{\ee}{\end{equation}}
\newcommand{\bes}{\begin{split}}
\newcommand{\es}{\end{split}}
\newcommand{\bee}{\begin{eqnarray}}
\newcommand{\eee}{\end{eqnarray}}
\newcommand{\beee}{\begin{array}}
\newcommand{\bem}{\begin{multline}}
\newcommand{\eem}{\end{multline}}
\newcommand{\bec}{\begin{Comment}}
\newcommand{\ec}{\end{Comment}}

\newcommand{\Y}[1]{{\ensuremath{\mathbf{Y}\{#1\}}}}
\newcommand{\Yy}{\ensuremath{\mathbf{Y}}}

\newcommand{\fm}[1]{_{\boldsymbol{{\scriptstyle #1}}}}

\newcommand{\aA}{{\ensuremath{\mathcal{A}}}}
\newcommand{\aB}{{\ensuremath{\mathcal{B}}}}

\newcommand{\auxiliary}{{\it auxiliary}}
\newcommand{\Stueckelberg}{{\it Stueckelberg}}
\newcommand{\dynamical}{{\it dynamical}}

\newcommand{\metric}{{\it metric-like}}
\newcommand{\constrained}{{\it constrained}}
\newcommand{\unconstrained}{{\it unconstrained}}
\newcommand{\lightcone}{{\it light-cone}}

\newcommand{\framelike}{{\it frame-like}}

\newcommand{\unfolded}{{\it unfolded}}

\newcommand{\WW}{{\ensuremath{\mathcal{W}}}}

\newcommand{\smallpic}[1]{{\unitlength=0.2mm#1}}

\newcommand{\DD}{{{\mathcal{D}}}}

\newcommand{\DL}{{D}}

\newcommand{\smallone}{\makebox[0pt][l]{$\scriptstyle 1$}}
\newcommand{\smalltwo}{\makebox[0pt][l]{$\scriptstyle 2$}}

\newcommand{\TypicalDiagramV}{{\bep(120,170){%
\put(0,20){\RectT{5}{4}{\TextTopLeftIn{\ensuremath{\scriptstyle s_N-1}}{\ensuremath{\scriptstyle p_N}}}}%
\put(0,60){\RectT{8}{4}{\TextTopLeftIn{\ensuremath{\scriptstyle s_2-1}}{\ensuremath{\scriptstyle p_2}}}}%
\put(0,100){\RectT{12}{5}{\TextTopLeftIn{\ensuremath{\scriptstyle s_1-1}}{\ensuremath{\scriptstyle p_1}}}}%
}\eep}}
\newcommand{\GeneralCase}{{\bep(120,140)(0,20){%
\put(0,0){\TypicalDiagramV}%
}\eep}}
\newcommand{\GeneralCaseA}{{\bep(120,140)(0,20){%
\put(0,0){\TypicalDiagramV}%
\put(0,10)\YoungA%
}\eep}}
\newcommand{\GeneralCaseB}{{\bep(120,140)(0,20){%
\put(0,0){\TypicalDiagramV}%
\put(0,10)\YoungB%
}\eep}}

\begin{document}
\renewcommand{\thefootnote}{\fnsymbol{footnote}}
{\begin{titlepage}
\begin{flushright}
\vspace{1mm}
FIAN/TD/19-08\\
\end{flushright}

\vspace{1cm}

\begin{center}
{\bf \Large Frame-like Actions for Massless Mixed-Symmetry Fields in Minkowski space} \vspace{1cm}

\textsc{E.D.
Skvortsov\footnote{skvortsov@lpi.ru}}

\vspace{.7cm}

{ I.E.Tamm Department of Theoretical Physics, P.N.Lebedev Physical
Institute,\\Leninsky prospect 53, 119991, Moscow, Russia}
\end{center}

\vspace{0.5cm}

\begin{abstract}
A frame-like action for arbitrary mixed-symmetry bosonic massless fields in Minkowski space is constructed. The action is given in a simple form and consists of two terms for a field of any spin. The fields and gauge parameters are certain tensor-valued differential forms. The formulation is based on the unfolded form of equations for mixed-symmetry fields.
\end{abstract}
\end{titlepage}
\renewcommand{\thefootnote}{\arabic{footnote}}
\setcounter{footnote}{0}
\section*{Introduction}

Due to the ever-growing interest in field theories in higher dimensions, the study of mixed-symmetry fields, i.e., those that correspond generally to neither symmetric nor antisymmetric tensor representations of the Wigner little algebra, was started from the paper \cite{Curtright:1980yk} and continued further in \cite{Aulakh:1986cb,Ouvry:1986dv,Labastida:1986gy,Labastida:1986ft,Labastida:1987kw}.

In $4d$ the spin is a single (half)integer. In the general case the spin degrees of freedom are in one-to-one correspondence with finite-dimensional unitary representations of the Wigner little algebra, which is $\mls$\footnote{Strictly speaking the stability subalgebra of a light-like momentum is $\isoLittle$, however $\isoLittle$-translations have to be realized trivially for massless fields, reducing $\isoLittle$ to $\mls$.} for massless fields and $\msv$ for massive ones. Therefore, the spin can be naturally characterized by Young diagrams. The most developed cases correspond to one-row (symmetric) and one-column (antisymmetric) diagrams (tensors), in the former case the $4d$-spin is equal to the number $n$ of boxes in one-row Young diagram ($n+1/2$ for fermions).

The theory of massless mixed-symmetry fields has been developing over the last decades within different approaches, which can be split into \lightcone, \metric\ and \framelike\ ones\footnote{Let us also mention the ambient approach of \cite{Metsaev:1995re}, see \cite{Fotopoulos:2008ka} for recent developments.}, in accordance with the types of fields used.

Within the \lightcone\ approach one deals with tensors of the Wigner little algebra and, hence, manifest Lorentz symmetry is lost. Nevertheless, the \lightcone\ approach turned out to be very effective in constructing cubic vertices, which is the first nontrivial attempt of introducing interactions. For instance, the cubic vertices of mixed-symmetry fields for $d\geq6$ were constructed in \cite{Metsaev:1993mj, Metsaev:1993ap} (see \cite{Metsaev:2005ar} for a review), inspiring the investigation to find full non-linear theories of mixed-symmetry fields.
The fist attempts towards an explicit construction of manifestly covariant cubic vertices for mixed-symmetry gauge fields were made in \cite{Boulanger:2004rx, Bekaert:2004dz}.

Within the \metric\ approach fields are the world tensors, which are analogous to metric field $g_{\mu\nu}$. The \metric\ approach can be split further into \constrained\ and \unconstrained\ approaches, according to whether or not some trace constraints are imposed on fields and gauge parameters\footnote{For the review of all approaches see \cite{Bekaert:2006ix}.}.

Within the \unconstrained\ approach of \cite{Francia:2002aa,Francia:2002pt,Bekaert:2002dt,Bekaert:2003az,Francia:2005bu}  fields and gauge parameters are not subjected to any trace constraints. A nice feature of the \unconstrained\ approach is its relation to the low-tension limit of free string field theory \cite{Francia:2002pt, Bonelli:2003kh, Sagnotti:2003qa, Francia:2005bu, Francia:2006hp}. However, the actions for unconstrained mixed-symmetry fields are only available in a nonlocal form \cite{Bekaert:2006ix}. Also, within the BRST approach, in which no off-shell trace constraints are imposed, the fields with the spin corresponding to two-row Young diagrams were studied in detail \cite{Burdik:2001hj, Burdik:2000kj, Buchbinder:2007ix, Moshin:2007jt}.

The \constrained\footnote{In principle, fields can be taken to be irreducible Lorentz tensors, though the gauge parameters in addition to being algebraically irreducible must satisfy differential constraints \cite{Skvortsov:2007kz}. } approach is ascribed to Fronsdal who showed in \cite{Fronsdal:1978rb} that totally symmetric massless fields have to be subjected to double-trace constraints for the equations to be gauge invariant and to describe the correct number of physical degrees of freedom \cite{Curtright:1979uz}. In the pioneering paper by Labastida \cite{Labastida:1987kw} on massless mixed-symmetry fields, a set of generalized trace constraints for fields and gauge parameters was suggested, the gauge invariant equations were derived and the local action was constructed. Unfortunately, it was not proved in \cite{Labastida:1987kw} that the correct number of physical degrees of freedom propagates, this was proved in \cite{Bekaert:2006ix}.

Making use of Fock space oscillators allowed Labastida to write down constraints/gauge transformations/equations/action for all mixed-symmetry fields at once. However, due to the lack of Young symmetry constraints a Fock space vector contains in its decomposition multiple copies of the same representations. In order to single out a particular mixed-symmetry field Young symmetry projectors are required. For individual mixed-symmetry fields the constraints/the field equations/the gauge transformations were obtained in \cite{Bekaert:2006ix}. To single out the action for a particular mixed-symmetry field from \cite{Labastida:1987kw} appears to be a very involved procedure.

Within the \framelike\ approach \cite{Vasiliev:1980as} fields are tensor-valued differential forms, which are generalizations of vielbein $e^a_\mu$ and Lorentz spin-connection $\omega^{a,b}_\mu$. The \framelike\ approach has a lot of advantages: the use of differential forms simplifies introducing interactions with gravity;  the forms take values in certain irreducible representations\footnote{The forms that take values in reducible representation of the Lorentz or (anti)-de Sitter algebra describe reducible sets of massless fields \cite{Sorokin:2008tf}, which are related to the tensionless limit of string theory \cite{Ouvry:1986dv, Bengtsson:1986ys, Pashnev:1989gm, Francia:2002pt, Sagnotti:2003qa}. This possibility can be referred to as the \unconstrained\ \framelike\ approach. } of the Lorentz \cite{Lopatin:1987hz, Zinoviev:2003ix, Zinoviev:2003dd} or (anti)-de Sitter algebra \cite{Vasiliev:2001wa, Alkalaev:2003qv, Alkalaev:2005kw, Alkalaev:2006rw, Skvortsov:2006at}, i.e., tangent tensors are not subjected to complicated double-trace constraints; frame-like fields can be recognized as Yang-Mills connections of the space-time symmetry algebra;  the most important is that frame-like fields can be embedded into the full set of fields of the \unfolded\ approach. From this perspective the \framelike\ approach is a very promising one.

The \unfolded\ approach itself is a reformulation of field equations in the first order form with the help of exterior differential and differential forms \cite{Vasiliev:1988xc, Vasiliev:1988sa, Vasiliev:1992gr}. The \unfolded\ approach appears to have a very rich underlying structure, so-called Free Differential Algebras \cite{Sullivan77}, which are also used in supergravity and M-theory \cite{D'Auria:1982nx, D'Auria:1982pm, Nieuwenhuizen:1982zf, Fre:2005px, Fre:2008qw}.

Among arbitrary-symmetry fields, distinguished is a subclass of totally-symmetric massless higher-spin fields, which has been the most extensively studied.
Free totally-symmetric higher-spin fields were reformulated within the \unfolded\ approach \cite{Vasiliev:1988xc} and then the consistent nonlinear deformation of the unfolded equations was found in $4d$ \cite{Vasiliev:1989yr, Vasiliev:1990en}, generalized further to arbitrary space-time dimension in \cite{Vasiliev:2003ev}.

The crucial ingredient is to find a higher-spin algebra, which is a non-abelian infinite-dimensional extension of the space-time symmetry algebra satisfying admissibility condition \cite{Konshtein:1988yg}, i.e. a gauging of the algebra must match some unitary representation thereof. The admissibility condition relates group- and field-theoretical aspects of the theory, stating that in order for some gauge theory with gauge algebra $\mathfrak{g}$ to be consistent the spectrum of relativistic fields that live on the solutions of the equations of motion has to match some unitary representation of $\mathfrak{g}$.

As for totally-symmetric higher-spin fields, the admissibility condition for higher-spin algebras turned out to be very restrictive, discarding most of possible higher-spin multiplets. For example, the simplest higher-spin algebra contains a single copy of a field of each spin $s=0,1,2,...$. For this reason, until admissible higher-spin algebras for mixed-symmetry fields are known it would be better to work with individual fields rather than fix any particular infinite multiplet.

Recently the unfolded form of equations for arbitrary mixed-symmetry bosonic and fermionic massless fields in the Minkowski space has been constructed \cite{Skvortsov:2008vs}. It turned out that the rather unnatural algebraic constraints that must be imposed both on fields and gauge parameters within the \constrained\ approach of Labastida can be easily explained. The Labastida fields and gauge parameters can be identified with certain components of tensor-valued differential forms, connections of the space-time symmetry algebra. It is worth stressing again that the \unfolded\ approach succeeded in constructing full non-linear equations for totally-symmetric higher-spin fields.

In this letter we construct a simple \framelike\ action for arbitrary-spin massless mixed-symmetry fields in Minkowski space.  In contrast to the \constrained\ approach, where the number of terms in
the action grows rapidly with the rank of a tensor, within the
\framelike\ approach the action consists only of two terms for an
arbitrary mixed-symmetry field; the Lagrangian equations are manifestly gauge
invariant.

In Section \ref{MMS} we recall the general properties of massless mixed-symmetry fields in Minkowski space,
essential features of the \unfolded\ approach are given in Section \ref{Unfld}. The action is constructed in Section \ref{MSActions}.

\section*{Notation}
Greek indices $\mu, \nu,...=0...(d-1)$ are the world indices of the Minkowski space-time. Lowercase Latin letters $a, b, c,...=0...(d-1)$ are the tangent indices in the basis that is defined by background vielbein $h^a_\mu$. $h^a_\mu h^b_\nu\eta_{ab}=g_{\mu\nu}$, $g_{\mu\nu}$ is a metric tensor in some coordinates and $\eta_{ab}=diag(1,-1,...,-1)$ is the invariant tensor of \lorentz. In Cartesian coordinates with $h^a_\mu=\delta^a_\mu$  there is no distinction between world and tangent tensors.

A group of $n$ (anti)symmetric indices $a_1...a_n$ or $ab...c$ is denoted ($a[n]$) $a(n)$ or ($[ab...c]$) $(ab...c)$. (Anti)symmetrization is denoted by placing a group of indices in (square)round brackets, or by designating indices by the same letter. Only necessary permutations are performed\footnote{Note that the (anti)symmetrization operation defined in this way is just an abridged notation for a number of distinct terms with permuted indices rather than a well-behaved projector.}, e.g., for a vector $V^a$ and a symmetric rank-two tensor $U^{ab}=U^{ba}$ $ V^{(a}U^{bc)}\equiv V^aU^{bc}+V^cU^{ab}+V^bU^{ca}$ and $V^aU^{aa}\equiv V^{a_1}U^{a_2a_3}+V^{a_2}U^{a_3a_1}+V^{a_3}U^{a_1a_2}$.

All the necessary information on Young diagrams and mixed-symmetry tensors is collected in Appendix \ref{AppYoungDiagrams}.

\section{Massless Mixed-Symmetry Fields}\setcounter{equation}{0}\label{MMS}

Though we work with world tensors in this Section, in order to simplify the comparison with the \unfolded\ approach, it is useful to convert all world tensor indices to tangent ones by virtue of inverse vielbein $h^{\mu a}$, e.g., $\pl^a\equiv h^{\mu a}\pl_\mu $, $\phi^{a(s)}\equiv\phi_{\mu_1...\mu_s}h^{\mu_1a}...h^{\mu_s a}$.

We start by briefly reviewing the general properties of totally-symmetric and totally-antisymmetric fields, whose spin degrees of freedom are characterized\footnote{The diagrams are given by specifying the lengths of the rows as $\Y{s_1,...,s_n}$ or by specifying the widths and the heights of its rectangular subblocks, e.g., $\Y{(s,1)}\equiv\Y{s}$ and $\Y{(1,p)}\equiv\Y{\underbrace{1,...,1}_p}$. See Appendix \ref{AppYoungDiagrams} for more detail.} by $\Y{(s,1)}$ and $\Y{(1,p)}$ irreducible representations of the Wigner massless little algebra \mls, respectively.

\textbf{A symmetric spin-$s$ field.} As was found by Fronsdal in \cite{Fronsdal:1978rb}, a totally-symmetric spin-$s$ massless field can be described\footnote{In fact, there is no strong interdependence between the representation $\Yy$ of \mls, i.e., the spin, and the representation of the Lorentz algebra $\Yy_M$ in which the field $\phi(x)$ takes values. The most important and the most natural choice is when $\Yy=\Yy_M$ as Young diagrams, all other are referred to as dual formulations. Note, that $\Yy$ corresponds to an irreducible representation of \mls, whereas $\Yy_M$ correspond to a representation of \lorentz\ that has to be reducible in most cases in order for equations of motion to be invariant under gauge transformations with differentially unconstrained parameters and for these field equations to admit a Lagrangian. For the minimal choice the term 'spin-$\Yy$ field' is unambiguous.} by a symmetric rank-$s$ tensor field $\phi^{(a_1...a_s)}$, satisfying
\be\label{FlatMSFronsdal}\begin{split}
    &\square\phi^{a_1...a_s}-\pl^{(a_1}\pl_m\phi^{m a_2...a_s)}+
    \pl^{(a_1}\pl^{a_2}\phi^{a_3...a_s) m}_{\phantom{a_3...a_s)m}m}=0,\quad
    \phi^{m\phantom{m}n\phantom{n}a_5...a_s }_{\phantom{m}m\phantom{n}n}\equiv0,\\
    &\delta
    \phi^{a_1...a_s}=\pl^{(a_1}\xi^{a_2...a_s)},\quad
    \xi^{m\phantom{m}a_3...a_{s-1}}_{\phantom{m}m}\equiv0,
\end{split}\ee
where in order for gauge parameter $\xi^{a_1...a_{s-1}}$ to be differentially unconstrained only the second trace of the field has to vanish, i.e., the field has to take values in a reducible representation of \lorentz.

\textbf{An antisymmetric ($p$-form) field.} Analogously, a totally-antisymmetric massless field or $p$-form can be described  by an antisymmetric rank-$p$ tensor field
$\omega^{[c_1...c_p]}$, satisfying
\be\square
\omega^{c_1...c_p}-\pl^{[c_1}\pl_{m}\omega^{m c_2...c_p]}=0,
\quad \delta\omega^{c_1...c_p}=
\pl^{[c_1}\xi^{c_2...c_p]},\ee
where gauge parameter $\xi^{c_1...c_{p-1}}$ is a rank-$(p-1)$ antisymmetric tensor. The gauge symmetry is reducible in the sense that not all of the gauge parameters do affect the field, these are represented by the second level gauge parameters $\xi^{c_1...c_{p-2}}$, $\delta \xi^{c_1...c_{p-1}}=\pl^{[c_1}\xi^{c_2...c_{p-1}]}$ modulo those components of $\xi^{c_1...c_{p-2}}$ that do not affect $\xi^{c_1...c_{p-1}}$ and so on until $\delta \xi^c=\pl^c \xi$.

Mixed-symmetry massless fields join together nontrivial trace constraints of symmetric fields with reducible gauge symmetries of $p$-form fields and introduce a new feature of having more than one gauge parameter. Let us consider the case of a spin-$\Y{s,t}$, i.e., spin-{\smallpic{\RectBRow{6}{3}{${\scriptstyle s}$}{${\scriptstyle t}$}}}, field in detail and review general properties of arbitrary-spin mixed-symmetry fields.

\textbf{A spin-$\Y{s,t}$ field.} A spin-$\Y{s,t}$ massless field can be described \cite{Labastida:1987kw} by a field $\phi^{a(s),b(t)}\equiv\phi^{a_1...a_s,b_1...b_t}$, which is symmetric in $a_1...a_s$ and $b_1...b_t$, separately, and $\phi^{(a_1...a_s,a_{s+1})b_2...b_{t}}\equiv0$, i.e., $\phi^{a(s),b(t)}$ has the symmetry of {\smallpic{\RectBRow{6}{3}{${\scriptstyle s}$}{${\scriptstyle t}$}}}. Analogously to symmetric fields, $\phi^{a(s),b(t)}$ is subjected to double-trace constraints with respect to each of the two groups of symmetric indices
\be\label{MSLabastidaDT}
\phi^{m\phantom{m}n\phantom{n}a_5...a_s,b_1...b_t}_{\phantom{m}m\phantom{n}n}\equiv0,\qquad \phi^{a_1...a_s,m\phantom{m}n\phantom{n}b_5...b_t}_{\phantom{a_1...a_s,m}m\phantom{n}n}\equiv0,
\ee
the cross traces with respect to a pair of indices from distinct groups need not vanish. Therefore, $\phi^{a(s),b(t)}$ contains a lot of irreducible components in general. The Labastida equations \cite{Labastida:1987kw}
\be\label{MSLabastida}\begin{split}
    \square &\phi^{a(s),b(t)}-\pl^a\pl_c\phi^{a(s-1)c,b(t)}-\pl^b\pl_c\phi^{a(s),b(t-1)c}+\\
    &+\pl^a\pl^b\phi^{a(s-1)c,b(t-1)}_{\phantom{a(s-1)c,b(t-1)}c}+
    \pl^a\pl^a\phi^{a(s-2)c\phantom{c},b(t)}_{\phantom{a(s-2)c}c}+
    \pl^b\pl^b\phi^{a(s),b(t-2)c}_{\phantom{a(s),b(t-2)c}c}=0,
\end{split}\ee
share the symmetry and the trace properties of $\phi^{a(s),b(t)}$. The equations are invariant under gauge transformations\footnote{Some work with Young symmetrizers is needed to extract the formulae for particular mixed-symmetry field from the results of \cite{Labastida:1987kw}. For example, the last term on the {\textit{r.h.s.}} of (\ref{MSLabastidaGauge}) supplements the second one for the whole expression to have the symmetry of  $\phi^{a(s),b(t)}$. In the general case, necessity for Young symmetrizers greatly complicates the issue of extracting given mixed-symmetry field out of \cite{Labastida:1987kw}. }
\be\label{MSLabastidaGauge}\delta\phi^{a(s),b(t)}={\frac{s-t}{s-t+1}}\pl^a\xi^{a(s-1),b(t)}_1-\pl^b\xi^{a(s),b(t-1)}_2+{\frac1{s-t+1}}\pl^a\xi_2^{a(s-1)b,b(t-1)},\ee
where gauge parameters $\xi^{a(s-1),b(t)}_1$ and $\xi^{a(s),b(t-1)}_2$ have the symmetry of {\smallpic{\RectBRow{5}{3}{${\scriptstyle s-1}$}{${\scriptstyle t}$}}} and {\smallpic{\RectBRow{6}{3}{${\scriptstyle s}$}{${\scriptstyle t-1}$}}}, respectively, and
\begin{align}
&\label{MSTwoRowTraceA} \xi^{a(s-3)m\phantom{m},b(t)}_{\smallone\phantom{a(s-3)m}m}+{\frac{2}{(s-t-1)}}\left(\xi^{a(s-3)bm,\phantom{m}b(t-1)}_{\smallone\phantom{a(s-3)bm,}m}+
{\frac{2}{(s-t)}}\xi^{a(s-3)bb,m\phantom{m}b(t-2)}_{\smallone\phantom{a(s-3)bb,m}m}\right)\equiv0,\\
&\label{MSTwoRowTraceB} \xi^{a(s),m\phantom{m}b(t-3)}_{\smalltwo\phantom{a(s),m}m}\equiv0,\\
&\label{MSTwoRowTraceC}\mbox{any double trace of } \xi^{a(s-1),b(t)}_1\ \mbox{and}\ \xi^{a(s),b(t-1)}_2\quad \mbox{vanishes},\\
&\label{MSTwoRowTraceD} {\frac{2(s-t)}{(s-t-1)}}\left(\xi^{a(s-2)m,\phantom{m}b(t-1)}_{\smallone\phantom{a(s-2)m,}m}+
{\frac{1}{(s-t+1)}}\xi^{a(s-2)b,m\phantom{m}b(t-2)}_{\smallone\phantom{a(s-2)b,m}m}\right)\equiv\xi^{a(s-2)m\phantom{m},b(t-1)}_{\smalltwo\phantom{a(s-2)m}m}+\nonumber\\
&\qquad+{\frac2{(s-t+1)}}\left(\xi^{a(s-2)bm,\phantom{m}b(t-2)}_{\smalltwo\phantom{a(s-2)bm,}m}+
{\frac2{(s-t+2)}}\xi^{a(s-2)bb,m\phantom{m}b(t-3)}_{\smalltwo\phantom{a(s-2)bb,m}m}\right),\\
&\label{MSTwoRowTraceE} \xi^{a(s-1),m\phantom{m}b(t-2)}_{\smallone\phantom{a(s-1),m}m}\equiv2\xi^{a(s-1)m,\phantom{m}b(t-1)}_{\smalltwo\phantom{a(s-1)m,}m}.
\end{align}
Identities (\ref{MSTwoRowTraceA}) and (\ref{MSTwoRowTraceB}) imply that the trace of $\xi^{a(s-1),b(t)}_1$ with the symmetry of {\smallpic{\RectBRow{5}{3}{${\scriptstyle s-3}$}{${\scriptstyle t}$}}} and the trace of $\xi^{a(s),b(t-1)}_2$ with the symmetry of {\smallpic{\RectBRow{6}{3}{${\scriptstyle s}$}{${\scriptstyle t-3}$}}} vanish. Both $\xi^{a(s-1),b(t)}_1$ and $\xi^{a(s),b(t-1)}_2$  have traces with the symmetry of {\smallpic{\RectBRow{5}{3}{${\scriptstyle s-2}$}{${\scriptstyle t-1}$}}} and {\smallpic{\RectBRow{5}{3}{${\scriptstyle s-1}$}{${\scriptstyle t-2}$}}}, (\ref{MSTwoRowTraceD}) and (\ref{MSTwoRowTraceE}) imply that these traces are not independent, being proportional to each other.

In contrast to both symmetric and antisymmetric fields there are two gauge parameters, whose Young diagrams can be obtained by cutting off one cell from the spin Young diagram {\smallpic{\RectBRow{6}{3}{${\scriptstyle s}$}{${\scriptstyle t}$}}} in various ways, i.e., {\smallpic{\RectBRow{5}{3}{${\scriptstyle s-1}$}{${\scriptstyle t}$}}} and {\smallpic{\RectBRow{6}{3}{${\scriptstyle s}$}{${\scriptstyle t-1}$}}}. However, by the construction the traces of $\xi^{a(s-1),b(t)}_1$ and $\xi^{a(s),b(t-1)}_2$ are not independent. This peculiarity of algebraic conditions will get a simple explanation within the \unfolded\ approach. Similarly to $p$-form fields the gauge symmetry is reducible
\begin{align}
    \delta \phi^{a(s),b(t)}=0&&\mbox{provided}&&\left\{\begin{array}{l}
    \delta\xi_1^{a(s-1),b(t)}=\pl^b\chi^{a(s-1),b(t-1)}-{\scriptstyle\frac1{s-t}}\pl^a\chi^{a(s-2)b,b(t-1)},\\
    \delta\xi_2^{a(s),b(t-1)}=\pl^a\chi^{a(s-1),b(t-1)}
     \end{array}\right.
\end{align}
where the second order gauge parameter $\chi^{a(s-1),b(t-1)}$ is traceless with respect to each pair of indices and has the symmetry of {\smallpic{\RectBRow{5}{3}{${\scriptstyle s-1}$}{${\scriptstyle t-1}$}}}, i.e., $\chi^{a(s-1),ab(t-2)}\equiv0$.

\textbf{A general mixed-symmetry field.} In the general case of a spin-$\Yy=\Y{s_1,...,s_n}$ massless field, field $\phi_\Yy$ is a Lorentz tensor with the symmetry of $\Yy$ and in addition to the Young symmetry constraints
$\phi^{{a}(s_1),{b}(s_2),...,d(s_n)}$ satisfies \cite{Labastida:1986ft}
\be\label{FlatMSLabastidaDoubleTracelessness}\eta_{cc}\eta_{dd}\phi^{a(s_1),...,b(s_i-4)ccdd,...,f(s_n)}\equiv0,\qquad
i\in[1,n],\ee
i.e., the second trace with respect to any four indices from the same group of symmetric indices must vanish. Gauge parameters at the $r$-th level of reducibility have the symmetry of
\be\label{FlatMSGaugeParameters}{\unitlength=0.3mm{\GaugeParameters}}:\quad\sum_{i=1}^{i=N}k_i=\sum_{i=1}^{i=N}p_i-r,\ee
where it is convenient to combine the rows of equal length into blocks, i.e., $\Yy=\Y{(s_1,p_1),...,(s_N,p_N)}$. Therefore, the first level gauge parameters are obtained by cutting off one cell from the bottom-right of any block. Evidently, there are $N$ gauge parameters at the first level. The trace conditions on the first level gauge parameters are more complicated than (\ref{FlatMSLabastidaDoubleTracelessness}) (cf. (\ref{MSTwoRowTraceA})-(\ref{MSTwoRowTraceE})), see \cite{Labastida:1987kw, Bekaert:2006ix} for more details, the general statement is that the traces of gauge parameters are not independent. This fact suggests that all gauge parameters can be incorporated into a single object as they do within the \unfolded\ approach.

\section{Essential of the Unfolded approach}\setcounter{equation}{0}\label{Unfld}
The \unfolded\ approach \cite{Vasiliev:1988xc, Vasiliev:1988sa, Vasiliev:1992gr} is a reformulation of field-theoretical systems as
\be\label{UnfldEquations} d W^\aA = F^\aA(W),\ee
where $W^\aA$ is a set of differential forms over the space-time, which do not necessary have the same degree, $q_\aA$ being the degree of $W^\aA$, $d$ - exterior differential on the space-time manifold and $F^\aA(W)$ is a degree-$(q_\aA+1)$ function that is assumed to be expandable in terms of wedge products only\footnote{Further the wedge symbol $\wedge$ will be systematically omitted.}, i.e.,
\be
F^\aA(W)=\sum_{n=1}^{\infty}\sum_{q_{\aB_1}+...+q_{\aB_n}=q_\aA+1}
f^\aA_{\phantom{\aA} \aB_1 ... \aB_n}W^{\aB_1}\wedge...\wedge
W^{\aB_n},\ee where $f^\aA_{\phantom{\aA} \aB_1 ... \aB_n}$ are
constant coefficients satisfying $f^\aA_{\phantom{\aA}
\aB_1...\aB_i\aB_j...\aB_n}=(-)^{q_{\aB_i}q_{\aB_j}}f^\aA_{\phantom{\aA}
\aB_1...\aB_j\aB_i...\aB_n}$. $F^\aA(W)$ is also assumed to satisfy an integrability condition, called generalized Jacobi identity,
\be\label{UnfldBianchi} F^\aB\frac{\vec{\pl}}{\pl
W^\aB}F^\aA\equiv0,\ee
which can be obtained by applying $d$ to (\ref{UnfldEquations}). Any solution of (\ref{UnfldBianchi}) is said to define a Free Differential Algebra  \cite{Sullivan77, D'Auria:1982nx,
D'Auria:1982pm, Nieuwenhuizen:1982zf}. In the field-theoretical sense (\ref{UnfldBianchi}) corresponds to Bianchi identities.

By virtue of (\ref{UnfldBianchi}), equations (\ref{UnfldEquations}) are invariant under gauge transformations
\begin{align}\label{UnfldGauge}&\delta
W^\aA=d\xi^\aA+\xi^\aB\frac{\vec{\pl}}{\pl
W^\aB}F^\aA,&&& \mbox{for }q_{\aA}>0,\\\label{UnfldGaugeZeroForm}&\delta
W^\aA=\xi^{\aB'}\frac{\vec{\pl}}{\pl
W^{\aB'}}F^\aA,& \aB': q_{\aB'}=1, & & \mbox{for
}q_{\aA}=0,\end{align}
where $\xi^\aA$ is a degree-$(q_\aA-1)$ form with values in the same space as $W^\aA$. Due to the fact that each $W^\aA$ of degree $q_\aA>0$ possesses its own gauge parameter $\xi^\aA$, nonzero degree fields are gauge fields. Provided that $F^\aA(W)$ is linear in a subset $\omega^i\subset W^\aA$, $\delta
\omega^i=0$ itself can be treated as unfolded system with respect to $\xi^i$ and, hence, the gauge transformations that does not affect the field $\omega^i$, i.e., the second level gauge transformations, are of the form
\be\label{UnfldGaugeRed}\delta
\xi^i=d\chi^i-\chi^j\frac{\vec{\pl}
}{\pl \omega^j}F^i,\ee
where $\chi^i$ is a degree-$(q_i-2)$ form, and so on for deeper levels gauge symmetries. Therefore, a degree-$q_i$ gauge field $\omega^i$ have exactly $q_i$ levels of gauge transformations.

Consequently, equations of motions, gauge symmetries, reducible gauge symmetries and Bianchi identities are the simple consequences of the single algebraic identity (\ref{UnfldBianchi}). For example, if we are given some gauge transformations the equations and the Bianchi identities can be easily recovered. In addition, the use of differential forms makes \unfolded\ approach manifestly covariant.

The most simple example of an unfolded system is given by one-form connection $\Omega^I_\mu$ of some Lie algebra $\mathfrak{g}$ with structure coefficients $f^I_{JK}$, satisfying
\be d\Omega^I=-f^I_{JK}\Omega^J\Omega^K.\label{UnfldFlatness}\ee
The generalized Jacobi identity reduces to the ordinary Jacobi identity on $f^I_{JK}$ and (\ref{UnfldFlatness}) is a zero Yang-Mills strength (flatness) condition on field $\Omega^I_\mu$. We take $\Omega^I_\mu$ to be a connection of the Poincare algebra, i.e., $\Omega^I=\{\varpi^{a,b}, h^a\}$, where $\varpi^{a,b}\equiv\varpi^{a,b}_\mu $ is a Lorentz spin-connection and $h^a\equiv h^a_\mu $ is a background vielbein. Any nontrivial solution ($h^a_\mu$ is a nonsingular matrix) of flatness equations
\begin{align}
        &dh^a+\varpi^{a,}_{\phantom{a,}b}h^b=0,\label{UnfldIsoFlatnessA}\\
        &d\varpi^{a,b}+\varpi^{a,}_{\phantom{a,}c}\varpi^{c,b}=0\label{UnfldIsoFlatnessB}
\end{align}
provides us with the basis of a tangent space $h^a_\mu$ and with Lorentz spin-connection $\varpi^{a,b}_\mu$, the latter is used to define a Lorentz covariant derivative of differential forms with values in any representation of \lorentz, i.e., having some tangent indices,
\be \DL T^{ab...}=dT^{ab...}+\varpi^{a,}_{\phantom{a,}c}T^{cb...}+\varpi^{b,}_{\phantom{b,}c}T^{ac...}+... \quad.\ee
A simple solution of (\ref{UnfldIsoFlatnessA}-\ref{UnfldIsoFlatnessB}) is given by Cartesian coordinates $h^a_\mu=\delta^a_\mu$, $\varpi^{a,b}_\mu=0$. It is assumed further that $h^a_\mu$ and $\varpi^{a,b}_\mu$ satisfy (\ref{UnfldIsoFlatnessA}-\ref{UnfldIsoFlatnessB}) but the advantage of the \unfolded\ approach is that no explicit solution is needed either to write down field-equations or to construct actions, being most effective in (anti)-de Sitter \cite{Lopatin:1987hz, Vasiliev:2001wa, Alkalaev:2003qv, Alkalaev:2005kw, Alkalaev:2006rw}.

The set of forms $W^\aA$ consists of a background connection $\Omega^I$ and dynamical fields $\omega^i$, i.e., $W^\aA=\{\{\varpi^{a,b}, h^a\},\omega^i\}$. As free fields only are considered, $h^a_\mu$ and $\varpi^{a,b}_\mu$ are assumed to be of order zero and $\omega^i$ are of the first order, with the equations on $\omega^i$ being linear in $\omega^i$. The dynamical fields $\omega^i$ are certain forms with values in finite-dimensional irreducible representations of \lorentz, i.e., just irreducible Lorentz tensors. The most general linear unfolded equations on $\omega^i$ are of the form
\be
\DL\omega^i=-\sum_j\sigma^{ij}(h,...,h,\omega^j)\label{UnfldGeneralLinear},
\ee
where the part of the Poincare connection associated with the Lorentz-spin connection $\varpi^{a,b}$ is combined with $d$ into the Lorentz covariant derivative $\DL$, which acts diagonally on $\omega^i$; $\sigma^{ij}$ are certain operators, which can be polynomial in $h^a$ and satisfy (\ref{UnfldBianchi}) since (\ref{UnfldIsoFlatnessA}-\ref{UnfldIsoFlatnessB}) are equivalent to $\DL h^a=0$ and $\DL^2=0$. No wonder that polynomial in vielbein $h^a_\mu$ can appear since field equations contain $g_{\mu\nu}g^{\lambda\rho}\phi_{\lambda\rho}$-like terms, which are polynomial in background metric $g_{\mu\nu}$. Free Differential Algebras are closely connected with Lie algebras, in fact, coefficients of $\sigma^{ij}$ correspond either to modules or to Chevalley-Eilenberg cocyles of the Poincare algebra \cite{Vasiliev:2007yc}.

It turns out \cite{Skvortsov:2008vs} that the unfolded equations for massless mixed-symmetry fields in Minkowski space are simpler than (\ref{UnfldGeneralLinear}) and have the form
\be \label{UnfldMSSimple} \DL\omega^{g}=-\sigma_-^g(h,...,h,\omega^{g+1}),\qquad g=0,1,..., \ee
where (\ref{UnfldBianchi}) reduces to $\sigma_-^g(h,...,h,\sigma_-^{g+1}(h,...h,\omega^{g+2}))\equiv0$ and the subscript ${}_-$ is due to the fact that all fields can be graded by non-negative integer $g$ and $\sigma_-$ is of grade $(-1)$. Dynamical field $\omega^g$ is a degree-$q_g$ form that takes values in irreducible representation $\Yy_g$ of \lorentz, i.e., the tangent tensor has the symmetry of Young diagram $\Yy_g$ and is completely traceless. The rank of tensor $\Yy_g$ is a strictly increasing function of $g$.  The operator $\sigma_-^g$ is uniquely determined by $\Yy_g$ and $\Yy_{g+1}$, it contracts each extra index of $\Yy_{g+1}$ as compared to $\Yy_g$ with $h^a$ and imposes, if needed, appropriate Young symmetrizer\footnote{For instance, let $\omega^0$ and $\omega^1$ be irreducible Lorentz tensors with the symmetry of $\smallpic{\YoungAA}$ and $\smallpic{\YoungBA}$, respectively, i.e., $\omega^0\equiv\omega^{a,b}=-\omega^{b,a}$ and $\omega^1\equiv\omega^{ab,c}=\omega^{ba,c}$ : $\omega^{(ab,c)}\equiv0$. Then,  $\DL\omega^{0}=\sigma_-(\omega^{1})$ with $\sigma_-(\omega^1)=h_c\omega^{ac,b}+\frac12h_c\omega^{ab,c}$, where the second term(Young symmetrizer) is added in order for the whole expression to have the symmetry of $\omega^0$.}. Therefore, $q_g$ is also completely determined by its value $q_0$ at $g=0$, $(q_g-q_{g+1}+1)$ being equal to the number of vielbeins to be contracted. Young diagrams $\Yy_g$ are determined by the spin $\Yy$ of the \metric\ field $\phi_\Yy(x)$.

If we denote $\sigma^g_-(h,..,h,\omega^{g+1})$ simply as $\sigma_-(\omega^{g+1})$, (\ref{UnfldBianchi}) is just a nilpotency of $\sigma_-$, $\sigma_-^2=0$. Technically, the nilpotency of $\sigma_-$ is due to the fact that when applied twice to any field, say\footnote{A degree-$q$ form with values in irreducible representation $\Yy$ of \lorentz, i.e., the tangent tensor has the symmetry of $\Yy$ and is traceless,  is denoted as $\omega^\Yy\fm{q}$.} $\omega^{\Yy_g}\fm{q_g}$, it contracts two vielbeins with the same group of symmetric indices, which is identically zero because of $h^ah^b=-h^bh^a$.

It is convenient to define the space of gauge fields $\WW\fm{q_0}=\{\omega^{\Yy_0}\fm{q_0}, \omega^{\Yy_1}\fm{q_1},...\}$, the space of the first level gauge parameters $\WW\fm{q_0-1}=\{\xi^{\Yy_0}\fm{q_0-1}, \xi^{\Yy_1}\fm{q_1-1},...\}$, which contains a finite number of parameters due to the fact that $q_g=0$ for sufficiently large $g$. Analogously, we can introduce the spaces $\WW\fm{q_0-k}$ of deeper levels gauge parameters and the space of field strengths $\WW\fm{q_0+1}=\{R^{\Yy_0}\fm{q_0+1}, R^{\Yy_1}\fm{q_1+1},...\}$, the spaces of Bianchi identities $\WW\fm{q_0+k}=\{B^{\Yy_0}\fm{q_0+k}, B^{\Yy_1}\fm{q_1+k},...\}$, $k>1$. With these definitions the unfolded equations and gauge transformations can be rewritten simply as
\begin{align}\label{ResultsFullSystem}
    \DD\omega\fm{p}&=0, &\qquad \omega\fm{p}&\in\WW\fm{p},\nonumber\\
    \delta \omega\fm{p}&=\DD\xi\fm{p-1},& \qquad \xi\fm{p-1}&\in\WW\fm{p-1},\nonumber\\
    \delta \xi\fm{p-1}&=\DD\xi\fm{p-2},&\qquad \xi\fm{p-2}&\in\WW\fm{p-2},\nonumber\\
    ..&..,& ..&..,\nonumber\\
    \delta \xi\fm{1}&=\DD\xi\fm{0}, &\qquad \xi\fm{0}&\in\WW\fm{0},
\end{align}
where $\DD=\DL+\sigma_-$ is a nilpotent operator $\DD^2=0$. The nilpotency of $\DD$ is a simple consequence of (i) nilpotency of  $\DL$, $(\DL)^2=0$ (\ref{UnfldIsoFlatnessB}); (ii) zero-torsion $\DL h^a=0$ (\ref{UnfldIsoFlatnessA}); (iii) $(\sigma_-)^2=0$. Field strength $R\fm{p+1}=\DD\omega\fm{p}$ satisfies Bianchi identity $B\fm{p+2}=\DD R\fm{p+1}\equiv0$.

Inasmuch as unfolded equations $\DD\omega\fm{p}=0$ can be treated \cite{Bekaert:2005vh} as a sort of cocycle condition $d\omega=0$, by virtue of the Poincare lemma it follows that all fields except for zero-degree forms are pure gauge and, hence, in order for unfolded equations to describe a field-theoretical system with propagating degrees of freedom, zero-forms have to be included.

\textbf{A spin-two field} \cite{Vasiliev:1986td, Vasiliev:1988xc, Vasiliev:1988sa, Vasiliev:2001wa, Vasiliev:2003ev}. To make the above statements more clear let us consider the example of a free massless spin-two field, i.e., the linearized gravitation. The field $\omega^{\Yy_0}\fm{q_0}$ is a well-known dynamical vielbein $e^a\fm{1}\equiv e^a_\mu dx^\mu$, the auxiliary field $\omega^{\Yy_1}\fm{q_1}$ is a dynamical Lorentz spin-connection $\omega^{ab}\fm{1}\equiv \omega^{ab}_\mu dx^\mu=-\omega^{ba}_\mu dx^\mu$, the first unfolded equation
\be \label{UnfldSpinTwoA}\DL e^{a}\fm{1}+h_c\omega^{ac}\fm{1}=0\ee
is invariant under gauge transformations
\be\label{UnfldSpinTwoB}\delta e^{a}\fm{1}=\DL\xi^{a}\fm{0}+h_c\xi^{ac}\fm{0},\qquad \delta\omega^{ab}\fm{1}=\DL\xi^{ab}\fm{0}, \ee
where $\xi^a\fm{0}\equiv \xi^a$ and $\xi^{ab}\fm{0}\equiv \xi^{ab}$ are the gauge parameters associated with $e^{a}\fm{1}$ and $\omega^{ab}\fm{1}$, respectively. We have to find $F^\aA(W)$ in the sector of $\omega^{ab}\fm{1}$ such that (i) linearized Einstein equations are imposed; (ii) Bianchi identities $h_c \DL\omega^{ac}\fm{1}\equiv0$ are satisfied. $\DL\omega^{ab}\fm{1}$ (in components $\frac12(D_\mu\omega^{ab}_\nu-D_\nu\omega^{ab}_\mu)$) is equal to the linearized Riemann tensor $R^{ab}_{\mu\nu}$, where it is useful to introduce its tangent version $R^{ab,cd}=R^{ab}_{\mu\nu}h^{\mu c}h^{\nu d}$.
Linearized Einstein equations $R_{\mu\nu}-\frac12 g_{\mu\nu}R_{\lambda\rho}g^{\lambda\rho}=0$ reduce to $R^{ac,bd}\eta_{cd}=0$, i.e., the trace of the Riemann tensor is zero but the traceless part of $R^{ab,cd}$, called Weyl tensor $C^{ab,cd}$, need not vanish. If we impose $\DL\omega^{ab}\fm{1}=0$, then, in addition to the linearized Einstein equations the Weyl tensor is also required to vanish, making the system dynamical empty. A Weyl tensor has the symmetry of {\smallpic{\YoungBB}}, i.e., it is antisymmetric in pairs $ab$, $cd$ and $C^{[ab,c]d}\equiv0$. Therefore, the correct unfolded equations have the form
\be \DL\omega^{ab}\fm{1}=-h_ch_dC^{ab,cd}\fm{0},\ee
where a zero-form $C^{ab,cd}\fm{0}$ can be recognized as a Weyl tensor, which is the only component of $R^{ab,cd}$ that can be nontrivial on-mass-shell.
Bianchi identities $h_b \DL\omega^{ab}\fm{1}=h_bh_ch_dC^{ab,cd}\equiv0$ are satisfied due to the Young symmetry of a Weyl tensor and the fact that vielbeins $h^a$ anti-commute. Again, $\DL C^{ab,cd}=0$ can not be imposed as it constraints the Weyl tensor to be a constant, and so on. The full unfolded system has the form \cite{Vasiliev:1988sa}
\begin{align}\label{UnfldSpiTwoFullsystem}
    &\DL e^{a}\fm{1}+h_c\omega^{ac}\fm{1}=0,&&\delta e^{a}\fm{1}=\DL\xi^{a}\fm{0}+h_c\xi^{ac}\fm{0},&&\nonumber\\
    &\DL\omega^{ab}\fm{1}+h_ch_dC^{ab,cd}\fm{0}=0, && \delta\omega^{ab}\fm{1}=\DL\xi^{ab}\fm{0},&&\\
    &\DL C^{aa,bb,c(k)}\fm{0}+h_dC^{aa,bb,c(k)d}\fm{0}=0, && \delta C^{aa,bb,c(k)}\fm{0}=0,&&k=0,1,..., \nonumber\end{align}
where a set of zero-forms $C^{aa,bb,c(k)}\fm{0}$ that are irreducible Lorentz tensors with the symmetry\footnote{$C^{aa,bb,c(k)}$ is antisymmetric in pairs $aa$, $bb$, symmetric in $c_1...c_k$, traceless with respect to each pair of indices and satisfies $C^{[aa,a]b,c(k)}\equiv0$, $C^{aa,[bb,b]c(k-1)}\equiv0$.} of \smallpic{\RectAYoung{5}{$k$}{\YoungBB}} is to be introduced.

The identification with the \metric\ approach is as follows. The field $e^{a|b}=e^a_\mu h^{\mu b}$ is to be decomposed as $e^{a|b}=\frac12\phi^{(ab)}+\frac12\psi^{[ab]}$, where $\psi^{[ab]}$ and $\phi^{(ab)}$ are an antisymmetric and a traceful symmetric fields, respectively. In terms of $\phi^{ab}$ and $\psi^{ab}$ gauge transformations (\ref{UnfldSpinTwoB}) read $\delta \phi^{ab}=\pl^a\xi^b+\pl^b\xi^a$, $\delta \psi^{ab}=\pl^a\xi^b-\pl^b\xi^a+2\xi^{ab}$. By virtue of a pure algebraic gauge symmetry with $\xi^{ab}$ field $\psi^{ab}$ can be gauged away. Therefore, the dynamical field is $\phi^{ab}$, which can be recognized as the Fronsdal field (\ref{FlatMSFronsdal}).

Since field equations are of the second order, (\ref{UnfldMSSimple}) at $g=0$ expresses the first auxiliary field in terms of the first derivatives of dynamical Labastida field $\phi_\Yy\in\omega^{\Yy_0}\fm{q_0}$, (\ref{UnfldMSSimple}) at $g=1$ imposes dynamical equations on $\phi_\Yy$ and expresses the second auxiliary field in terms of the second derivatives of $\phi_\Yy$, higher equations ($g>1$) impose no nontrivial constraints on $\phi_\Yy$. Further, we concentrate on the first two unfolded equations only, since it is sufficient for constructing Lagrangians.

\textbf{A spin-$s$ field} \cite{Vasiliev:1986td, Vasiliev:1988xc, Vasiliev:1988sa, Vasiliev:2001wa, Vasiliev:2003ev}. The above can be generalized to a totally-symmetric spin-$s$ field. In accordance with the general construction of \cite{Skvortsov:2008vs},  $\Yy_0={\smallpic{\RectARow{5}{$\scriptstyle s-1$}}}$, $\Yy_1={\smallpic{\RectBRow{5}{1}{$\scriptstyle s-1$}{}}}$, $\Yy_2={\smallpic{\RectBYoung{5}{$\scriptstyle s-1$}{\YoungB}}}$ and $q_0=q_1=q_2=1$. Therefore, the dynamical field $\phi^{a(s)}$ is incorporated into a frame-like one-form $e^{a(s-1)}\fm{1}\equiv e^{a(s-1)}_\mu dx^\mu$, the first auxiliary field is $\omega^{a(s-1),b}\fm{1}$ and the first two unfolded equations together with the gauge transformations have the form
\begin{align}\label{UnfldHSFullsystem}
    &\DL e^{a(s-1)}\fm{1}+h_c\omega^{a(s-1),c}\fm{1}=0,&&  \delta e^{a(s-1)}\fm{1}=\DL \xi^{a(s-1)}\fm{0}+h_c\xi^{a(s-1),c}\fm{0},\nonumber\\
    &\DL \omega^{a(s-1),b}\fm{1}+h_c\omega^{a(s-1),bc}\fm{1}=0,&&  \delta \omega^{a(s-1),b}\fm{1}=\DL \xi^{a(s-1),b}\fm{0}+h_c\xi^{a(s-1),bc}\fm{0}.
\end{align}
Similarly to the spin-two case $e^{a(s-1)|b}=e^{a(s-1)}_\mu h^{\mu b}$ decomposes into two fields $e^{a(s-1)|b}=\frac1s\phi^{a(s-1)b}+\frac1s\psi^{a(s-1),b}$ with the symmetry of ${\smallpic{\RectARow{5}{$\scriptstyle s$}}}$ and ${\smallpic{\RectBRow{5}{1}{$\scriptstyle s-1$}{}}}$, respectively. Note that $e^{a(s-1)|b}$ is traceless with respect to $a_1,...,a_{s-1}$ and, hence, the second trace of $\phi^{a(s)}$ vanishes. In terms of $\phi$ and $\psi$ the gauge transformations have the form  $\delta\phi^{a(s)}=\pl^a\xi^{a(s-1)}$ and $\delta\psi^{a(s-1),b}=(s-1)\pl^b\xi^{a(s-1)}-\pl^a\xi^{a(s-2)b}+s\xi^{a(s-1),b}$. Therefore, the traceless part of $\psi^{a(s-1),b}$ can be gauged away\footnote{Note that the trace of the Fronsdal field $\phi^{a(s)}$ is also contained in $\psi^{a(s-1),b}$, namely, $2\phi^{a(s-2)m}_{\phantom{a(s-2)m}m}=(s-2)\psi^{a(s-2)m,}_{\phantom{a(s-2)m,}m}$. Nevertheless, the dynamical components of $e^{a(s-1)}\fm{1}$ are equivalent to the Fronsdal field $\phi^{a(s)}$.} by virtue of $\xi^{a(s-1),b}$. With world indices converted to the tangent ones according to $\omega^{a(s-1),b|c}\equiv\omega^{a(s-1),b}_\mu h^{c\mu}$, $\omega^{a(s-1),bb|c}\equiv\omega^{a(s-1),bb}_\mu h^{c\mu}$, the first two unfolded equations (\ref{UnfldHSFullsystem}) are rewritten as
\begin{align}
\label{UnfldHSEqA}&\pl^ce^{a(s-1)|d}-\pl^d e^{a(s-1)|c}=\omega^{a(s-1),c|d}-\omega^{a(s-1),d|c},\\
\label{UnfldHSEqB}&\pl^c\omega^{a(s-1),b|d}-\pl^d\omega^{a(s-1),b|c}=\omega^{a(s-1),bc|d}-\omega^{a(s-1),bd|c}.
\end{align}
Symmetrizing $c$ with $a_1...a_{s-1}$ in (\ref{UnfldHSEqA}),  (\ref{UnfldHSEqB}) and, then, contracting (\ref{UnfldHSEqB}) with $\eta_{bd}$ results in
\begin{align}
&\omega^{a(s-1),d|a}=\pl^de^{a(s-1)|a}-\pl^ae^{a(s-1)|d},\label{UnfldTSA}\\
&\omega^{a(s-1),m|}_{\phantom{a(s-1),m|}m}=\pl_m e^{a(s-2)m|a}-\pl^a e^{a(s-2)m|}_{\phantom{a(s-2)m|}m},\label{UnfldTSB}\\
&\pl_m\omega^{a(s-1),m|a}-\pl^a\omega^{a(s-1),m|}_{\phantom{a(s-1),m|}m}=0,\label{UnfldTSC}
\end{align}
where (\ref{UnfldTSB}) is a contraction of (\ref{UnfldTSA}) with $\eta_{da_{s}}$. Note that $\omega^{a(s-1),bc}\fm{1}$ does note contribute to the dynamical equations of motion since it is auxiliary. Substituting (\ref{UnfldTSA}), (\ref{UnfldTSB}) into (\ref{UnfldTSC}) and, then, making use of $\phi^{a(s)}=e^{a(s-1)|a}$ and $\phi^{m\phantom{m}a(s-2)}_{\phantom{m}m}=2e^{a(s-2)m|}_{\phantom{a(s-2)m|}m}$ gives the Fronsdal equations (\ref{FlatMSFronsdal}) for $\phi^{a(s)}$.

\textbf{A spin-$\Y{s,t}$ field}. According to the general recipe of \cite{Skvortsov:2008vs} $\Yy_0={\smallpic{\RectBRow{6}{4}{$\scriptstyle s-1$}{$\scriptstyle t-1$}}}$, $\Yy_1={\smallpic{\RectCYoung{6}{4}{{$\scriptstyle s-1$}}{{$\scriptstyle t-1$}}{\YoungA}}}$, $\Yy_2={\smallpic{\RectCYoung{6}{4}{{$\scriptstyle s-1$}}{{$\scriptstyle t-1$}}{\YoungB}}}$ and $q_0=q_1=q_2=2$. The Labastida field $\phi^{a(s),b(t)}$ is to be incorporated into the frame-like field $e^{a(s-1),b(t-1)}\fm{2}$ and an analog of Lorentz spin-connection is given by $\omega^{a(s-1),b(t-1),c}\fm{2}$. The first two unfolded equations
\begin{align}\label{UnfldTwoRowA}
    &R^{a(s-1),b(t-1)}\fm{3}\equiv\DL e^{a(s-1),b(t-1)}\fm{2}+h_d\omega^{a(s-1),b(t-1),d}\fm{2}=0,\\
    &R^{a(s-1),b(t-1),c}\fm{3}\equiv\DL \omega^{a(s-1),b(t-1),c}\fm{2}+h_d\omega^{a(s-1),b(t-1),cd}\fm{2}=0,\label{UnfldTwoRowB}
\end{align}
are obviously invariant under the first level gauge transformations
\begin{align}\label{UnfldTwoRowAA}
\delta e^{a(s-1),b(t-1)}&=\DL \xi^{a(s-1),b(t-1)}\fm{1}+h_d\xi^{a(s-1),b(t-1),d}\fm{1},\\
\delta \omega^{a(s-1),b(t-1),c}&=\DL \xi^{a(s-1),b(t-1),c}\fm{1}+h_d\xi^{a(s-1),b(t-1),cd}\fm{1},\\
\delta \omega^{a(s-1),b(t-1),cc}&=\DL \xi^{a(s-1),b(t-1),cc}\fm{1}+h_d\xi^{a(s-1),b(t-1),ccd}\fm{1},\label{UnfldTwoRowAAC}
\end{align}
where the last term in (\ref{UnfldTwoRowAAC}) is related to higher-grade fields and is not needed in what follows, the only point being that it has $\sigma_-(...)$ form.

In its turn the gauge fields are invariant under the second level gauge transformations
\begin{align}\label{UnfldTwoRowBB}
\delta \xi^{a(s-1),b(t-1)}\fm{1}&=\DL \chi^{a(s-1),b(t-1)}\fm{0}+h_d\chi^{a(s-1),b(t-1),d}\fm{0}\ ,\\
\delta \xi^{a(s-1),b(t-1),c}\fm{1}&=\DL \chi^{a(s-1),b(t-1),c}\fm{0}+h_d\chi^{a(s-1),b(t-1),cd}\fm{0},\\
\delta \xi^{a(s-1),b(t-1),cc}\fm{1}&=\DL \chi^{a(s-1),b(t-1),cc}\fm{0}+h_d\chi^{a(s-1),b(t-1),ccd}\fm{0}.
\end{align}
Let us draw the reader's attention to the conformity of equations with gauge transformations.
Note that gauge fields/parameters take values in irreducible representations of the Lorentz algebra, i.e., the tangent tensors have definite Young symmetry and are completely traceless. $e^{a(s-1),b(t-1)|cd}=e^{a(s-1),b(t-1)}_{\mu\nu}h^{\mu c}h^{\nu d}$ can easily be decomposed into irreducible tensors of \lorentz\ as
\begin{align}
{\smallpic{\RectBRow{6}{3}{$\scriptstyle s-1$}{$\scriptstyle t-1$}}}\otimes{\smallpic{\YoungAA}}&=\left[{\smallpic{\RectBRow{7}{4}{$\scriptstyle s$}{$\scriptstyle t$}}}\oplus{\smallpic{\RectBRow{6}{4}{$\scriptstyle s-1$}{$\scriptstyle t-1$}}}\oplus{\smallpic{\RectBRow{7}{3}{$\scriptstyle s$}{$\scriptstyle t-2$}}}\oplus{\smallpic{\RectBRow{5}{4}{$\scriptstyle s-2$}{$\scriptstyle t$}}}\oplus{\smallpic{\RectBRow{5}{3}{$\scriptstyle s-2$}{$\scriptstyle t-2$}}}\right]\oplus\nonumber\\
&\oplus\left[{\smallpic{\RectCYoung{7}{3}{{$\scriptstyle s$}}{{$\scriptstyle t-1$}}{\YoungA}}}\oplus{\smallpic{\RectCYoung{6}{3}{{$\scriptstyle s-1$}}{{$\scriptstyle t$}}{\YoungA}}}\oplus{\smallpic{\RectCYoung{6}{3}{{$\scriptstyle s-1$}}{{$\scriptstyle t-1$}}{\YoungAA}}}\oplus{\smallpic{\RectCYoung{5}{3}{{$\scriptstyle s-2$}}{{$\scriptstyle t-1$}}{\YoungA}}}\oplus{\smallpic{\RectCYoung{6}{3}{{$\scriptstyle s-1$}}{{$\scriptstyle t-2$}}{\YoungA}}}\oplus{\smallpic{\RectBRow{6}{3}{$\scriptstyle s-1$}{$\scriptstyle t-1$}}}\right],\label{UnfldTwoRowContent}
\end{align}
where all Young diagrams correspond to traceless tensors. The irreducible tensors in the first brackets of (\ref{UnfldTwoRowContent}) are the components of the Labastida field $\phi^{a(s),b(t)}=e^{a(s-1),b(t-1)|ab}$. Indeed, by virtue of the second level gauge parameter $\chi^{a(s-1),b(t-1),cd}\fm{0}$ with the symmetry of {\smallpic{\RectCYoung{6}{3}{{$\scriptstyle s-1$}}{{$\scriptstyle t-1$}}{\YoungB}}} the respective component of the first order gauge parameter $\xi^{a(s-1),b(t-1),c|d}=\xi^{a(s-1),b(t-1),c}_\mu h^{\mu d}$ can be gauged away, with the rest of irreducible components of $\xi^{a(s-1),b(t-1),c|d}$ exactly matching the terms in the second brackets of (\ref{UnfldTwoRowContent}). Therefore, it is $\phi^{a(s),b(t)}=e^{a(s-1),b(t-1)|ab}$ that can not be gauged away by an algebraic gauge symmetry, thus being a dynamical field. Note that there are two irreducible components with the symmetry of {\smallpic{\RectBRow{6}{3}{$\scriptstyle s-1$}{$\scriptstyle t-1$}}}, with one belonging to $\phi^{a(s),b(t)}$ and the other one being pure gauge.

The Labastida double-tracelessness (\ref{MSLabastidaDT}) is a consequence of the irreducibility of tangent tensors and of the lack of any trace conditions with respect to world and tangent indices. The Labastida gauge parameters $\xi_1^{a(s-1),b(t)}$ and $\xi_2^{a(s),b(t-1)}$ can be identified as
\begin{align}
 &\xi_1^{a(s-1),b(t)}=\xi^{a(s-1),b(t-1)|b}+{\scriptstyle\frac1{s-t}}\xi^{a(s-2)b,b(t-1)|a},\\
 &\xi_2^{a(s),b(t-1)}=\xi^{a(s-1),b(t-1)|a},
\end{align}
then, nontrivial trace constraints (\ref{MSTwoRowTraceA}-\ref{MSTwoRowTraceE}) are the consequence of this definition. To obtain, if needed at all, the Labastida equations (\ref{MSLabastida}) from the unfolded ones (\ref{UnfldTwoRowA}-\ref{UnfldTwoRowB}) is a simple 'symmetrizing and taking traces' problem.

\textbf{From \unfolded\ to \metric\ for a spin-$\Y{s,t}$ field.} We work in Cartesian coordinates, i.e., $h^a_\mu=\delta^a_\mu$, $\varpi^{a,b}_\mu=0$ and, hence, there is no distinction between world and tangent indices. With world indices converted to the tangent ones, the first two unfolded equations (\ref{UnfldTwoRowA})-(\ref{UnfldTwoRowB}) read
\begin{align}
\label{AppTwoRowA}&\pl^{[c}e^{a(s-1),b(t-1)|cc]}=-\omega^{a(s-1),b(t-1),[c|cc]},\\
\label{AppTwoRowB}&\pl^{[c}\omega^{a(s-1),b(t-1),d|cc]}=-\omega^{a(s-1),b(t-1),d[c|cc]},
\end{align}
Symmetrizing $c_1$, $c_2$ with $a_1...a_{s-1}$ and $b_1...b_{t-1}$ in (\ref{AppTwoRowA}-\ref{AppTwoRowB}), and taking trace with respect to $d$ and $c$ in (\ref{AppTwoRowB}) results in
\begin{align}
\label{AppTwoRowAA}&\pl^{a}e^{a(s-1),b(t-1)|bc}+\pl^{b}e^{a(s-1),b(t-1)|ca}+\pl^{c}e^{a(s-1),b(t-1)|ab}=-\omega^{a(s-1),b(t-1),c|ab},\\
\label{AppTwoRowBB}&\pl^{a}\omega^{a(s-1),b(t-1),m|b}_{\phantom{a(s-1),b(t-1),m|b}m}+
\pl^{b}\omega^{a(s-1),b(t-1),m|\phantom{m}a}_{\phantom{a(s-1),b(t-1),m|}m}+\pl_{m}\omega^{a(s-1),b(t-1),m|ab}=0.
\end{align}
Note that field $\omega^{a(s-1),b(t-1),cc}\fm{2}$ is auxiliary and does not contribute to the dynamical equations, being excluded by the projector onto the Labastida equations. The third term in (\ref{AppTwoRowBB}) can be directly expressed from (\ref{AppTwoRowAA}); to express the first two terms of (\ref{AppTwoRowBB}) we take the traces in (\ref{AppTwoRowAA}) with respect to $a_s$, $c$ and $b_t$, $c$
\begin{align}\label{AppTwoRowAAA}
&\pl^{a}e^{a(s-2)m,b(t-1)|b}_{\phantom{a(s-2)m,b(t-1)|b}m}+\pl^{b}e^{a(s-2)m,b(t-1)|\phantom{m}a}_{\phantom{a(s-2)m,b(t-1)|}m}+\pl_{m}e^{a(s-2)m,b(t-1)|ab}=-\omega^{a(s-1),b(t-1),m|\phantom{m}b}_{\phantom{a(s-1),b(t-1),m|}m},\\
&\pl^{a}e^{a(s-1),b(t-2)m|b}_{\phantom{a(s-1),b(t-2)m|b}m}+\pl^{b}e^{a(s-1),b(t-2)m|\phantom{m}a}_{\phantom{a(s-1),b(t-2)m|}m}+\pl_{m}e^{a(s-1),b(t-2)m|ab}=-\omega^{a(s-1),b(t-1),m|a\phantom{m}}_{\phantom{a(s-1),b(t-1),m|a}m}.\label{AppTwoRowBBB}
\end{align}

The Labastida field $\phi^{a(s),b(t)}$ is to be identified as $\phi^{a(s),b(t)}=e^{a(s-1),b(t-1)|ab}$, its various traces are given by
\begin{align}
&\phi^{a(s-2)m\phantom{m},b(t)}_{\phantom{a(s-2)m}m}=2e^{a(s-2)m,b(t-1)|\phantom{m}b}_{\phantom{a(s-2)m,b(t-1)|}m}, \qquad\phi^{a(s),b(t-2)m}_{\phantom{a(s),b(t-2)m}m}=2e^{a(s-1),b(t-2)m|a}_{\phantom{a(s-1),b(t-2)m|a}m},\nonumber\\
&\phi^{a(s-1)m,\phantom{m}b(t-1)}_{\phantom{a(s-1)m,}m}=e^{a(s-2)m,b(t-1)|a}_{\phantom{a(s-2)m,b(t-1)|a}m}+e^{a(s-1),b(t-2)m|\phantom{m}b}_{\phantom{a(s-1),b(t-2)|m}m}.\label{AppTwoRowIden}
\end{align}
Substituting (\ref{AppTwoRowAAA}), (\ref{AppTwoRowBBB}), (\ref{AppTwoRowAA}) into (\ref{AppTwoRowBB}) and, then, using (\ref{AppTwoRowIden}) results in the Labastida equations (\ref{MSLabastida}).

For the further comparison with the Lagrangian equations let us note that the dynamical equations have been obtained by making use of $R^{a(s-1),b(t-1)|abc}$ and $R^{a(s-1),b(t-1),m|ab}_{\phantom{a(s-1),b(t-1),m|ab}m}$ components of field strengths (\ref{UnfldTwoRowA}) and (\ref{UnfldTwoRowA}), instead of the whole field strengths. The auxiliary field $\omega^{a(s-1),b(t-1),cc}\fm{2}$ does not contribute to $R^{a(s-1),b(t-1),m|ab}_{\phantom{a(s-1),b(t-1),m|ab}m}$ by the construction.

It is significant that the above computations can be directly generalized to an arbitrary-spin mixed-symmetry field, resulting in the Labastida equations \cite{Labastida:1987kw}.

\textbf{A spin-$\Yy$ field}. The main statement of \cite{Skvortsov:2008vs} is that a spin-$\Yy$\footnote{$\Yy=\Y{(s_1,p_1),...,(s_N,p_N)}$ is taken in block's notation or $\Yy=\{h_1,h_2,...h_{s_1}\}$, $h_i$ being the height of the $i$-th column. Let $p=p_1+...+p_N$ be the height of $\Yy$. It is  obvious that $h_1=p$, $h_2=p$ if $s_N>1$ and $h_2=p-p_N$, otherwise.} massless field can be uniquely described within the \unfolded\ approach. $\Yy_0$ is obtained by cutting off the first column of $\Yy$, i.e., $\Yy_0=\Y{(s_1-1,p_1),...,(s_N-1,p_N)}=\Y{h_2,h_3,...,h_{s_1}}$, $\Yy_1=\Y{h_1+1,h_3,...,h_{s_1}}$,  $\Yy_2=\Y{h_1+1,h_2+1,h_4,...,h_{s_1}}$. $q_0=p=h_1$, $q_1=h_2$, $q_2=h_3$. For instance, when $s_N>2$
\be\label{UnfldDiagrams}\Yy_0=\smallpic{\GeneralCase}\quad\Yy_1=\smallpic{\GeneralCaseA}\quad\Yy_2=\smallpic{\GeneralCaseB}\ee
The sketch of the proof
\begin{enumerate}
  \item In order for gauge transformations to have $p$ levels, the Labastida field $\phi_\Yy$ has to be incorporated into certain degree-$p$ form $e^{\Yy_0}\fm{p}$. The gauge parameter at the $p$-th level is a degree-zero form $\xi^{\Yy_0}\fm{0}$ and, hence, $\Yy_0$ has the symmetry of the single gauge parameter at the level-$p$ (cf. (\ref{FlatMSGaugeParameters})). Converting all world form indices to the tangent ones
      \be\label{MSFrametoMetric}e^{a_1(s_1-1),...,a_p(s_p-1)|[d_1...d_p]}=e^{a_1(s_1-1),...,a_p(s_p-1)}_{\mu_1...\mu_p}h^{\mu_1d_1}...h^{\mu_pd_p},\ee
      the field $\phi_{\Yy}(x)$ is to be identified with $e^{a_1(s_1-1),...,a_p(s_p-1)|a_1...a_p}$. The Labastida double-tracelessness condition is a simple consequence of the tracelessness of the tangent tensor. All level-$k$ gauge parameters are contained in a single object $\xi^{\Yy_0}\fm{p-k}$, explaining the fact that the traces of gauge parameters are not independent within the \metric\ approach.

  \item The gauge symmetry was made manifest by the price of introducing redundant components that are given by various components of $e^{a_1(s_1-1),...,a_p(s_p-1)|[d_1...d_p]}$ that do not belong to $\phi_\Yy(x)$, gauge parameters may also have redundant components. In order to make redundant fields non-dynamical, an algebraic (\Stueckelberg) symmetry is introduced. Fortunately, all redundant components can be compensated by an algebraic gauge symmetry with a single $\xi^{\Yy_1}\fm{q_1-1}$
      \be\delta e^{\Yy_0}\fm{p}=\DL\xi^{\Yy_0}\fm{p-1}+\sigma_-(\xi^{\Yy_1}\fm{q_1-1}),\ee
      analogously for gauge parameters at the level-$k$
      \be\delta \xi^{\Yy_0}\fm{p-k}=\DL\xi^{\Yy_0}\fm{p-k-1}+\sigma_-(\xi^{\Yy_1}\fm{q_1-k-1}).\ee
  \item Since gauge parameter $\xi^{\Yy_1}\fm{q_1-1}$ is associated with the gauge field $\omega^{\Yy_1}\fm{q_1}$, the first unfolded equation is
      \be \DL e^{\Yy_0}\fm{p}+\sigma_-(\omega^{\Yy_1}\fm{q_1})=0.\ee
  \item The Bianchi identities $\sigma_-(\DL\omega^{\Yy_1}\fm{q_1})\equiv0$ can be solved \cite{Skvortsov:2008vs} as
      \be \label{UnfldGeneralSecondEq}\DL \omega^{\Yy_1}\fm{q_1}+\sigma_-(\omega^{\Yy_2}\fm{q_2})=0.\ee
  \item Equation (\ref{UnfldGeneralSecondEq}) implies new Bianchi identities and so on, resulting in the full unfolded system \cite{Skvortsov:2008vs}.
  \item Despite the unambiguity of unfolding, the facts that (i) correct second order equations are indeed imposed on the dynamical field $\phi_\Yy\in e^{\Yy_0}\fm{p}$; (ii) the rest of the unfolded equations imposes no additional differential constraints on $\phi_\Yy$; (iii) there are no other dynamical fields in the system; (iv) equations imposed indeed describe the correct number of physical degrees of freedom\footnote{This problem was not solved by Labastida in general, but was solved in \cite{Bekaert:2006ix}. As it has been already noted the spin-$\Y{s,t}$ example can be easily extended to the general case in order to derive the Labastida equations from the unfolded ones for arbitrary-spin field. This provides us with another consistency check of the Labastida work. }; have to be checked. The $\sigma_-$ cohomology technique \cite{Lopatin:1987hz,Shaynkman:2000ts} turned out to be very effective, solving all four problems at once \cite{Skvortsov:2008vs}.
\end{enumerate}

\section{Local Actions}\setcounter{equation}{0}\label{MSActions}
To begin with, let us note that the Maxwell action for a spin-one massless field  $S=-\frac14\int d^d x
F_{\mu\nu}F^{\mu\nu}$, where $F_{\mu\nu}=\pl_\mu A_\nu-\pl_\nu
A_\mu$ and $A_\mu$ is a potential, can be rewritten in the first order form \be \label{ActionMaxwellFirstOrder}S=\int d^d x
\left(\pl_\mu A_\nu-\pl_\nu
A_\mu+C_{\mu\nu}\right)C^{\mu\nu},\ee where
$C_{\mu\nu}$ is a rank-two antisymmetric auxiliary field. The equations for $C_{\mu\nu}$ are algebraic with respect to $C_{\mu\nu}$ and
can be easily solved as $C_{\mu\nu}=-\frac12F_{\mu\nu}$. The key moment is that (\ref{ActionMaxwellFirstOrder}) admits a reformulation in terms of differential forms, i.e., frame-like fields to be embedded into unfolded systems \be
S=\int \left(dA\fm{1}+\frac12h_bh_c
C^{bc}\right)C^{aa}\epsilon_{aav(d-2)}h^v...h^v,\ee where $A\fm{1}\equiv A_\mu$ is a Maxwell gauge potential one-form and $C^{ab}=-C^{ba}$ is a degree-zero form, which is antisymmetric in tangent indices and is a tangent version of $C_{\mu\nu}$. $\epsilon_{a_1...a_d}$ is a totally antisymmetric tensor, the Levi-Civita symbol. Use is made of
\be\underbrace{\epsilon_{a_1...a_d}\Xi^{a_1...a_k}\fm{k}h^{a_{k+1}}...h^{a_{d}}}_{\mbox{degree}-{\displaystyle{d}}\ \mbox{volume form}}=\left[\begin{tabular}{c}\mbox{Cartesian}\\ coordinates\end{tabular}\right]=\Xi^{a_1...a_k}_{\mu_1...\mu_k}\delta^{\mu[k]}_{a[k]},\ee
where $\Xi^{a_1...a_k}\fm{k}$ is a $k$-from with $k$ antisymmetric tangent indices and $\delta^{\mu[k]}_{a[k]}=\delta^{[\mu_1}_{a_1}...\delta^{\mu_k]}_{a_k}$.
A frame-like action for a totally-symmetric spin-$(s>1)$ field was constructed
in \cite{Vasiliev:1980as} \be\label{ActionSymmetric} S= \int(de^{u a(s-2)} +\frac{1}{2}h_c\,
\omega^{ua(s-2),c})\, \omega_{a(s-2)}{}^{u,\, u}\,\epsilon_{uuu
v(d-3)}\,h^v...h^v, \ee which operates with the first two fields, the dynamical one and the first auxiliary one,
of unfolded system (\ref{UnfldHSFullsystem}) and correctly
reproduces the dynamical equations. Action (\ref{ActionSymmetric}) is equivalent
to the Fronsdal one after solving the algebraic equations for auxiliary
field $\omega^{a(s-1),b}_\mu$ and expressing the terms with
$e^{a(s-1)}_\mu$ via $\phi_{\mu_1...\mu_s}$.

It is the action (\ref{ActionSymmetric}) that will be generalized to the action for an arbitrary mixed-symmetry field. Very instructive was the observation of \cite{Zinoviev:2003ix, Zinoviev:2003dd}, made on the basis of the simplest mixed-symmetry fields, that the indices of fields can be formally split into world and tangent ones.

Since dynamical equations are of the second order, it is sufficient to
make use of gauge fields/field strengths at grade zero and one\footnote{Recall that for a spin-$\Yy$ field, where
$\Yy=\Y{(s_1,p_1),...,(s_N,p_N)}=\Y{h_1,h_2...,h_{s_1}}$, $p=p_1+p_2+...+p_N$ is the height
of the first column of $\Yy$;  We abbreviate $q_1$ and $q_2$ as
$q$ and $r$, $q_0$ being equal to $p$ by definition. Note that
$q=p=h_2$ unless $s_N=1$ and $q=p-p_N=h_2$ otherwise. Then, $\Yy=\Y{p,q,r,h_4,...,h_{s_1}}$ $\Yy_0=\Y{q,r,h_4,...,h_{s_1}}$, $\Yy_1=\Y{p+1,r,h_4,...,h_{s_1}}$, $\Yy_2=\Y{p+1,q+1,h_4,...,h_{s_1}}$.}
\begin{align}\label{ActionFieldStrengthA}
R^0\fm{p+1}&=\DL e^{\Yy_0}\fm{p}+\sigma_-(\omega^{\Yy_1}\fm{q}), & &R^0\fm{p+1}\in \WW^{g=0}\fm{p+1},\quad e^{\Yy_0}\fm{p}\in \WW^{g=0}\fm{p} \\
\label{ActionFieldStrengthB}R^1\fm{q+1}&=\DL\omega^{\Yy_1}\fm{q}+\sigma_-(\omega^{\Yy_2}\fm{r}),
& &R^1\fm{q+1}\in \WW^{g=1}\fm{p+1},\quad \omega^{\Yy_1}\fm{q}\in
\WW^{g=1}\fm{p}, \quad \omega^{\Yy_2}\fm{r}\in \WW^{g=2}\fm{p}.
\end{align}

To construct an action, a degree-$d$ volume form that is  bilinear in fields has to be found. It is convenient to introduce for any $k=0...d$ a degree-$(d-k)$ form with $k$ antisymmetric indices $E_{u[k]}\equiv\epsilon_{u[k]b_1...b_{d-k}}h^{b_1}...h^{b_{d-k}}$, built with background vielbeins, which satisfies\footnote{(\ref{MSIdentityVielbeins}) can be derived from the identity $\epsilon_{[u_1...u_k b_2...b_{d-k+1}}\delta^c_{b_1]}h^{b_1}h^{b_2}...h^{b_{d-k}}\equiv0$.}
\be \label{MSIdentityVielbeins}h^c E_{u_1...u_k }=\frac1{d-k+1}\sum_{i=1}^{i=k}(-)^{i+k}\delta^c_{u_i}E_{u_1...\hat{u_i}...u_k}.\ee

For a degree-$(p'+1)$ form $\Phi^{\Yy_0}\fm{p'+1}$ with tangent indices
$a(s_1-1),b(s_2-1),...,c(s_r-1),d,...,e$ and a degree-$q'$ form
$\Psi^{\Yy_1}\fm{q'}$ with tangent indices $a(s_1-1),b(s_2-1),...,c(s_r-1),d,e,..,f$ provided that $p+q=p'+q'$ let us define a
scalar product
\be\label{MSActionScalarProduct}\langle\Phi\fm{p'+1}^{\Yy_0}|\Psi\fm{q'}^{\Yy_1}\rangle=\int
{\Phi\fm{p'+1}}^{ua(s_1-2),...,uc(s_r-2),\overbrace{\scriptstyle u,...,u}^{q-r}}{\Psi\fm{q'}}^{u\phantom{a(s_1-2)},...,u\phantom{c(s_r-2)},\overbrace{\scriptstyle u,u,...,u}^{p+1-r}}_{\phantom{u}a(s_1-2)\phantom{,...,}\phantom{u}c(s_r-2)}E_{u[p+q+1]}.\ee
The first index from each of the groups is contracted with $E_{u...u}$. Since $\Yy_0$ and $\Yy_1$ coincides modulo first column, the rest of tangent indices of $\Phi^{\Yy_0}\fm{p'+1}$ are contracted with the corresponding tangent indices of $\Psi^{\Yy_1}\fm{q'}$.

Taking into account that the sum of the heights of the first columns of $\Yy_0$ and $\Yy_1$ is equal to the
total degree $(p+q+1)$ of $\Phi\fm{p'+1}\Psi\fm{q'}$ and the tangent tensors have definite Young symmetry and are completely traceless, it is easy to see that the scalar
product possesses two important properties
\begin{align}&\langle\sigma_-(\Theta^{\Yy_1}\fm{q})|\Psi^{\Yy_1}\fm{q}\rangle=\langle\sigma_-(\Psi^{\Yy_1}\fm{q})|\Theta^{\Yy_1}\fm{q}\rangle,&&
\mbox{for any}\ \Theta^{\Yy_1}\fm{q}, \Psi^{\Yy_1}\fm{q} \in \WW^{g=1}\fm{p}\label{MSScalarPropertiesA},\\
&\langle\Phi\fm{p'}^{\Yy_0}|\sigma_-(\Upsilon\fm{r'}^{\Yy_2})\rangle=0,&&\mbox{for
any}\ \substack{\displaystyle \Phi^{\Yy_0}\fm{p'}\in\WW^{g=0}\fm{p'},\\\displaystyle \Upsilon^{\Yy_2}\fm{r'}\in
\WW^{g=2}\fm{2p-p'},}\ p'+r'=p+r.\label{MSScalarPropertiesB}\end{align}

\paragraph{The proof of (\ref{MSScalarPropertiesA}).} Let $\Theta\fm{q}^{\Yy_1}$ and $\Psi\fm{q}^{\Yy_1}$ be elements of $\WW^{g=1}\fm{p}$. Assume that $q=p$. Applying (\ref{MSIdentityVielbeins}) one obtains
\begin{align}
&h_m{\Theta\fm{q}}^{ua(s_1-2),...,uc(s_r-2),\overbrace{\scriptstyle u,...,u}^{q-r},m}
{\Psi\fm{q}}^{u\phantom{a(s_1-2)},...,u\phantom{c(s_r-2)},\overbrace{\scriptstyle u,...,u}^{p+1-r}}_{\phantom{u}a(s_1-2)\phantom{,...,}\phantom{u}c(s_r-2)\phantom{u}}
E_{u[p+1+q]}\sim\nonumber\\
&\sim\sum_{i=1}^{i=p+1}(-)^i{\Theta\fm{q}}^{ua(s_1-2),...,ub(s_i-2),...,uc(s_r-2),\overbrace{\scriptstyle u,...,u}^{p-r},m}
{\Psi\fm{q}}^{u\phantom{a(s_1-2)},...,\phantom{mb(s_i-2)},...,u\phantom{c(s_r-2)},\overbrace{\scriptstyle u,...,u}^{p+1-r}}_{\phantom{u}a(s_1-2)\phantom{,...,}mb(s_i-2)\phantom{,...,}\phantom{u}c(s_r-2)\phantom{u}}E_{u[2q]}\sim\nonumber\\
&\sim\sum_{i=1}^{i=p+1}(-)^{i}{\Theta\fm{q}}^{ua(s_1-2),...,b(s_i-1),...,uc(s_r-2),\overbrace{\scriptstyle u,...,u}^{p+1-r}}
{\Psi\fm{q}}^{u\phantom{a(s_1-2)},...,\phantom{b(s_i-1)},...,u\phantom{c(s_r-2)},\overbrace{\scriptstyle u,...,u}^{p+1-r}}_{\phantom{u}a(s_1-2)\phantom{,...,}b(s_i-1)\phantom{,...,}\phantom{u}c(s_r-2)\phantom{u}}E_{u[2q]},\nonumber
\end{align}
where the last expression is obviously symmetric with respect to $\Theta^{\Yy_1}\fm{q}$ and $\Psi^{\Yy_1}\fm{q}$, and
use is made of the following simple consequence of the Young symmetry properties\be C^{a(k_1),...,b(k_i-1)u,...,bc(k_j-1),...}=-C^{a(k_1),...,b(k_i),...,uc(k_j-1),...}\label{MSYoungConsequence}\ee
For the case $p>q$, $\sigma_-(\Theta^{\Yy_1}\fm{q})$ involves $(p+1-q)$ vielbeins \be h_{m_1}...h_{m_{p+1-q}}{\Theta\fm{q}}^{ua(s_1-2),...,uc(s_r-2),\overbrace{\scriptstyle u,...,u}^{q-r},m_1,...,m_{p+1-q}}.\ee
by applying (\ref{MSIdentityVielbeins}) $m_1,...,m_{p+1-q}$ turn out to be contracted with certain $u,...,u$ from the ${i_1}$,...,${i_{p+1-q}}$ groups of ${\Psi\fm{p}}^{\Yy_1}$, the sum over different rearrangements is implied.
If $i_k$ is in the range $1,...,r$ then $m_{i_k}$ appears to be symmetrized with $s_{i_k}-2$ indices of the $i_k$ group and, hence, by virtue of (\ref{MSYoungConsequence}) $m_{i_k}$ can be exchanged with $u_{i_k}$, the resulting expression being symmetric in $\Theta$ and $\Psi$.
If $i_k$ is in the range $r+1,...,p+1$  $m_{i_k}$ can be replaced with $u_{i_k}$ directly since $\Theta$ is explicitly antisymmetric in the indices from the groups $r+1,...,p+1$ (each group consists only of one index), and hence these terms can also be cast into the form that is explicitly symmetric with respect to $\Theta$ and $\Psi$. $\blacksquare$

\paragraph{The proof of (\ref{MSScalarPropertiesB}).} Let $\Phi\fm{p'}^{\Yy_0}$ and $\Upsilon^{\Yy_2}\fm{r'}$ be elements of $\WW^{g=0}\fm{p'}$ and $\WW^{g=2}\fm{2p-p'}$, $p+r=p'+r'$. Assume that $q=r$. Applying (\ref{MSIdentityVielbeins}) one obtains
\begin{align}
&{\Phi\fm{p'}}^{ua(s_1-2),...,uc(s_r-2)}
h_m{\Upsilon\fm{r'}}^{u\phantom{a(s_1-2)},...,u\phantom{c(s_r-2)},um,\overbrace{\scriptstyle u,...,u}^{p-q}}_{\phantom{u}a(s_1-2)\phantom{,...,u}c(s_r-2)}E_{u[p+q+1]}\sim\nonumber\\
&\sim\sum_{i=1}^{i=r}(-)^i{\Phi\fm{p'}}^{ua(s_1-2),...,mb(s_i-2),...,uc(s_r-2)}
{\Upsilon\fm{r'}}^{u\phantom{a(s_1-2)},...,u\phantom{b(s_i-2)},...,u\phantom{c(s_r-2)},u\phantom{m},\overbrace{\scriptstyle u,...,u}^{p-q}}
_{\phantom{u}a(s_1-2)\phantom{,...,u}b(s_i-2)\phantom{,...,u}c(s_r-2),\phantom{u}m}E_{u[p+q]}\sim\nonumber\\
&\sim\sum_{i=1}^{i=r}(-)^i{\Phi\fm{p'}}^{ua(s_1-2),...,b(s_i-1),...,uc(s_r-2)}
{\Upsilon\fm{r'}}^{u\phantom{a(s_1-2)},...,\phantom{b(s_i-1)},...,u\phantom{c(s_r-2)},uu,\overbrace{\scriptstyle u,...,u}^{p-q}}
_{\phantom{u}a(s_1-2)\phantom{,...,}b(s_i-1)\phantom{,...,u}c(s_r-2)}E_{u[p+q]}=0\nonumber
\end{align}
to obtain the last expression, which is identically zero since two antisymmetrized indices $u$ appear in the same group of symmetric indices, use was made of (\ref{MSYoungConsequence}). The extension on $q>r$ is similar to the proof of (\ref{MSScalarPropertiesA}):
among $h_{m_1}...h_{m_{q-r+1}}$ of
\be \sigma_-(\Upsilon^{\Yy_2}\fm{r'})\longleftrightarrow h_{m_1}...h_{m_{q-r+1}}{\Upsilon\fm{r'}}^{ua(s_1-2),...,u c(s_r-2),\overbrace{\scriptstyle um,...,um}^{q+1-r},\overbrace{\scriptstyle u,...,u}^{p-q}}\ee
at least one of $m$ will be contracted with some of the first $r$ groups of indices of $\Phi$ and hence by virtue of (\ref{MSYoungConsequence}) it can be exchanged with $u$, resulting in zero. $\blacksquare$

In order to derive the most general form of the action we use the three crucial observations: (\emph{a}) (\ref{MSActionScalarProduct}) is the only way to build a
degree-$d$ volume form of the elements of $\WW^{g=0}\fm{p+1}$ and $\WW^{g=1}\fm{p}$; (\emph{b}) there are two relevant elements of $\WW^{g=0}\fm{p+1}$, i.e., $\DL e^{\Yy_0}\fm{p}$ and $\sigma_-(\omega^{\Yy_1}\fm{q})$; (\emph{c}) there is only one relevant element of $\WW^{g=1}\fm{p}$, i.e., $\omega^{\Yy_1}\fm{q}$; the most general action is proved to be of
the form \be S=\langle
\DL e^{\Yy_0}\fm{p}+(1+\alpha)\sigma_-(\omega^{\Yy_1}\fm{q})|\,\omega^{\Yy_1}\fm{q}\rangle,\ee
with $\alpha$ being a free coefficient.

Lagrangian equations are to be of the form \begin{align}\label{ActionEqA}
\frac{\delta
S}{\delta\omega^{\Yy_1}\fm{q}}&=\pi_1[R^0\fm{p+1}]=0, \\ \label{ActionEqB}
\frac{\delta S}{\delta
e^{\Yy_0}\fm{p}}&=\pi_0[R^1\fm{q+1}]=\pi_0[\DL\omega^{\Yy_1}\fm{q}]=0,
\end{align}
where $\pi_1$ is a projector into $\omega^{\Yy_1}\fm{q}$ and $\pi_0$
is a projector into $e^{\Yy_0}\fm{p}$, i.e., $\pi_0$ and $\pi_1$ are the projectors induced by the contraction of $R^1\fm{q+1}$ and $R^0\fm{p+1}$ with $e^{\Yy_0}\fm{p}$ and $\omega^{\Yy_1}\fm{q}$, respectively. Since (i) all \dynamical\
fields are contained in $e^{\Yy_0}\fm{p}$, i.e., $\omega^{\Yy_1}\fm{q}$ is an
\auxiliary\ field; (ii) $R^0\fm{p+1}=0$ does not impose any dynamical equations on
$e^{\Yy_0}\fm{p}$ \cite{Skvortsov:2008vs}, expressing $\omega^{\Yy_1}\fm{q}$ in terms of the
first derivatives of $e^{\Yy_0}\fm{p}$ modulo $\sigma_-$ closed
terms; it follows that $\pi_1$ is trivial in the sense that it allows one to express non-exact part of
$\omega^{\Yy_1}\fm{q}$ via the first derivatives of $e^{\Yy_0}\fm{p}$. Since $R^1\fm{q+1}$ has two indices more than
$e^{\Yy_0}\fm{p}$, $\pi_0$ contracts two indices, one form and one
tangent, and imposes, if needed, the Young symmetry conditions,
which are trivial in symmetric basis. By virtue of (\ref{MSScalarPropertiesB}) $\pi_0$ sends
$\sigma_-(\omega^{\Yy_2}\fm{r})$ to zero in accordance with the fact that
$\omega^{\Yy_2}\fm{r}$ does not contribute to the dynamical
equations, i.e. the term $\langle e^{\Yy_0}\fm{p}|\sigma_-(\omega^{\Yy_2}\fm{r})\rangle$ can in principle be added to the action but it is identically zero.

The action is to be invariant under the standard gauge
transformations (\ref{ResultsFullSystem})
\begin{align}\delta e^{\Yy_0}\fm{p}&=\DL \xi^{\Yy_0}\fm{p-1}+\sigma_-(\xi^{\Yy_1}\fm{q-1}),& &\xi^{\Yy_0}\fm{p-1}\in \WW^{g=0}\fm{p-1},\quad \xi^{\Yy_1}\fm{q-1}\in \WW^{g=1}\fm{q-1},\\
\delta
\omega^{\Yy_1}\fm{q}&=\DL \xi^{\Yy_1}\fm{q-1}+\sigma_-(\xi^{\Yy_2}\fm{r-1}), && \xi^{\Yy_2}\fm{r-1}\in \WW^{g=2}\fm{r-1}\label{MSActionGaugeB}.\end{align}
Despite the fact that $\omega^{\Yy_2}\fm{r}$ does not contribute to the dynamical equations, the action reveals a symmetry with gauge parameter $\xi^{\Yy_2}\fm{r-1}$.

To check this invariance it is useful to rewrite the action as
$S=\langle
R^0\fm{p+1}|\omega^{\Yy_1}\fm{q}\rangle+\alpha\langle\sigma_-(\omega^{\Yy_1}\fm{q})|\omega^{\Yy_1}\fm{q}\rangle$.
The action turns out to be
invariant under gauge transformations both with
$\xi^{\Yy_0}\fm{p-1}$ and $\xi^{\Yy_2}\fm{r-1}$ for any $\alpha$. However in order to cancel the variation with respect to $\xi^{\Yy_1}\fm{q-1}$ the value of $\alpha=-\frac12$ is required. Indeed,
\be\delta S=\langle
R^0\fm{p+1}|\delta\omega^{\Yy_1}\fm{q}\rangle+2\alpha\langle\sigma_-(\omega^{\Yy_1}\fm{q})|\delta\omega^{\Yy_1}\fm{q}\rangle=\langle
R^0\fm{p+1}+2\alpha\sigma_-(\omega^{\Yy_1}\fm{q})|\delta\omega^{\Yy_1}\fm{q}\rangle,\ee where the symmetry property (\ref{MSScalarPropertiesA}) and $\delta R^0\fm{p+1}\equiv0$ have been used. Then, substituting (\ref{MSActionGaugeB})\be\delta S=\langle
R^0\fm{p+1}+2\alpha\sigma_-(\omega^{\Yy_1}\fm{q})|\DL \xi^{\Yy_1}\fm{q-1}\rangle+\langle
R^0\fm{p+1}+2\alpha\sigma_-(\omega^{\Yy_1}\fm{q})|\sigma_-(\xi^{\Yy_2}\fm{r-1})\rangle,\ee the second term vanishes due to (\ref{MSScalarPropertiesB}). By virtue of $\DL^2=0$ (\ref{UnfldIsoFlatnessB}) and $\DL h^a=0$ (\ref{UnfldIsoFlatnessA}) the $d$-form in the second term of \be\delta S=(1+2\alpha)\langle
\sigma_-(\omega^{\Yy_1}\fm{q})|\DL \xi^{\Yy_1}\fm{q-1}\rangle+\langle\DL e^{\Yy_0}\fm{p}
|\DL \xi^{\Yy_1}\fm{q-1}\rangle\ee is exact and, hence, $\delta S=0$ provided $\alpha=-\frac12$. Consequently, the action becomes \be \label{MSLocalAction}{S=\langle
\DL e^{\Yy_0}\fm{p}+{\textstyle\frac12}\sigma_-(\omega^{\Yy_1}\fm{q})|\,\omega^{\Yy_1}\fm{q}\rangle}\ee
Inasmuch as action (\ref{MSLocalAction}) is gauge invariant, the Lagrangian equations can be written in terms of field strengths (\ref{ActionFieldStrengthA}) and (\ref{ActionFieldStrengthB}). In Cartesian coordinates the variation of the action reads
\begin{align}
{\frac{\delta S}{\delta \omega}}\,\delta \omega=&{R_{\mu[p+1]}}^{ua(s_1-2),...,uc(s_r-2),\overbrace{\scriptstyle u,...,u}^{q-r}}
\delta{\omega_{\mu[q]}}^{u\phantom{a(s_1-2)},...,u\phantom{c(s_r-2)},\overbrace{\scriptstyle u,...,u}^{p+1-r}}_{\phantom{u}a(s_1-2)\phantom{,...,u}c(s_r-2)}\delta^{\mu[p+q+1]}_{u[p+q+1]}=0\Rightarrow \nonumber\\
&\Rightarrow R^{ua(s_1-2),...,uc(s_r-2),\overbrace{\scriptstyle u,...,u}^{q-r}|v[p+1]}
\delta\omega_{va(s_1-2),...,vc(s_r-2),\underbrace{\scriptstyle v,...,v}_{p+1-r}|u[q]}=0\Rightarrow\nonumber\\
&\Rightarrow R^{a_1(s_1-1),...,a_r(s_r-1),a_{r+1},...,a_{q}|a_1...a_r...a_q...a_{p+1}}=0\label{MSEquationsA}\\
{\frac{\delta S}{\delta e}}\,\delta e=&{R_{\mu[q+1]}}^{ua(s_1-2),...,uc(s_r-2),\overbrace{\scriptstyle u,...,u}^{p+1-r}}
\delta{e_{\mu[p]}}^{u\phantom{a(s_1-2)},...,u\phantom{c(s_r-2)},\overbrace{\scriptstyle u,...,u}^{q-r}}_{\phantom{u}a(s_1-2)\phantom{,...,u}c(s_r-2)}\delta^{\mu[p+q+1]}_{u[p+q+1]}=0\Rightarrow \nonumber\\ &\Rightarrow R^{ua(s_1-2),...,uc(s_r-2),\overbrace{\scriptstyle u,...,u}^{p-r},m|v[q]}_{\phantom{ua(s_1-2),...,uc(s_q-2),u,...,u,m|v[q]}m}
\delta e_{va(s_1-2),...,vc(s_r-2),\underbrace{\scriptstyle v,...,v}_{q-r}|u[p]}=0\Rightarrow\nonumber\\
&\Rightarrow R^{a_1(s_1-1),...,a_r(s_r-1),a_{r+1},...,a_q,...,a_p,m|a_1...a_q}_{\phantom{a_1(s_1-1),...,a_r(s_r-1),a_{r+1},...,a_q,...,a_p,m|a_1...a_q}m}=0\label{MSEquationsB}
\end{align}
which gives $\pi_1$ and $\pi_0$ with required properties. Indeed, after fixing the algebraic gauge symmetry for $e^{\Yy_0}\fm{p}$ the dynamical components of $e^{\Yy_0}\fm{p}$ are equivalent to the Labastida field $\phi_\Yy$. Then, by definition of $\Yy_1$ and $\sigma_-$, the nonzero components of $\DL e^{\Yy_0}\fm{p}$ match $\sigma_-$-nonclosed components of $\omega^{\Yy_1}\fm{q}$ (actually all $\sigma_-$-closed components of $\omega^{\Yy_1}\fm{q}$ are $\sigma_-$-exact and, hence, are pure gauge by virtue of $\xi^{\Yy_2}_{r-1}$ \cite{Skvortsov:2008vs}).

Note, that to derive the dynamical equations it is sufficient to set to zero the components of the field strengths $R^{\Yy_0}\fm{p+1}$ and $R^{\Yy_1}\fm{q+1}$ with the symmetry of $\Yy_d=\Y{h_1+1,h_2,...h_{s_1}}\equiv\Y{(s_1,p_1),...(s_N,p_N),1}$ and $\Yy$, respectively (see a spin-$\Y{s,t}$ example). It is the projection onto $\Yy_d$ and $\Yy$ that is made in (\ref{MSEquationsA}) and (\ref{MSEquationsB}). Obviously, $\pi_1$ allows one to express the components of $\omega^{\Yy_1}\fm{q}$ with the symmetry of $\Yy_d$ in terms of $\DL e^{\Yy_0}\fm{p}$ from $\pi_1 \left[R^{\Yy_0}\fm{p+1}\right]=0$. $\pi_0$ is even more trivial and allows one to project $R^{\Yy_1}\fm{q+1}$ onto components with the symmetry of $\Yy$.

For example, for a spin-\smallpic{\YoungBA} field action (\ref{MSLocalAction}) has the form \be
S=\int
\left[de^{u}\fm{2}+\frac12h_bh_c\omega^{ubc}\fm{1}\right]\omega^{uuu}\fm{1}\epsilon_{uuuuv(d-4)}h^v...h^v,\ee
and after passing to world indices coincides with \cite{Zinoviev:2003ix}. For a spin-{\smallpic{\RectBRow{6}{3}{${\scriptstyle s}$}{${\scriptstyle t}$}}} field action (\ref{MSLocalAction}) has the form
\be
S=\int
\left[d{e\fm{2}}^{ua(s-2),ub(t-2)}+\frac12h_c{\omega\fm{2}}^{ua(s-2),ub(t-2),c}\right]{\omega\fm{2}}^{u\phantom{a(s-2)},u\phantom{b(t-2)},u}_{\phantom{u}a(s-2)\phantom{,u}b(t-2)}\epsilon_{u(5)v(d-5)}h^v...h^v\ee
with the Lagrangian equations
\begin{align}
&R^{ua(s-2),ub(t-2)|mmm}\delta \omega_{ma(s-2),mb(t-2),m|uu}=0&&\Rightarrow&&
R^{a(s-1),b(t-1)|abc}=0,\\
&R^{ua(s-2),ub(t-2),c|\phantom{c}mm}_{\phantom{ua(s-2),ub(t-2),c|}c}\delta e_{ma(s-2),mb(t-2)|uu}=0&&\Rightarrow&&
R^{a(s-1),b(t-1),c|\phantom{c}ab}_{\phantom{a(s-1),b(t-1),c|}c}=0.
\end{align}
It is from these equations that the dynamical equations have been recovered in the previous section.

To conclude, we note that despite the fact that the action (\ref{MSLocalAction}) is not built of field strengths and, hence, is not manifestly gauge invariant, equations of motion (\ref{ActionEqA}-\ref{ActionEqB}) are written in terms of gauge invariant field strengths. Simplicity of the \framelike\ action (\ref{MSLocalAction}) is very encouraging and we hope it will be very helpful for the further study of mixed-symmetry fields.

\section*{Acknowledgements}
The author appreciates sincerely M.A.Vasiliev and K.B.Alkalaev for very valuable comments on a preliminary version of this work. The author is grateful to R.R.Metsaev and O.V.Shaynkman for many helpful and stimulating discussions while writing this paper.
The work was supported in part by
grants RFBR No. 08-02-00963, LSS-1615.2008.2, INTAS No. 05-7928, by the Landau scholarship and by the scholarship of the Dynasty foundation.

\appendix
\renewcommand{\theequation}{\Alph{section}.\arabic{equation}}

\section{Young diagrams}
\setcounter{section}{1} \label{AppYoungDiagrams}
\setcounter{equation}{0}

When working with mixed-symmetry tensors an essential use is made of Young diagrams. A Young diagram is a picture consisting of
$n$ left-justified rows made of boxes, with the $i$-th row containing $s_i$ boxes, $s_i$ being non-increasing function of $i$.
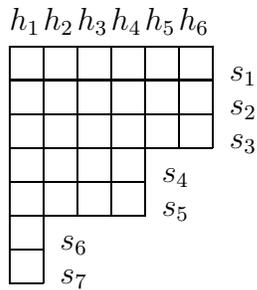
\begin{tabular}{p{3.4cm}p{11cm}} \ArbitraryYoungDiagram & There is a number of ways to define a particular Young diagram that are used in applications. A Young diagram can be defined by directly enumerating the lengths of rows $\Y{s_1,s_2,...,s_n}$, e.g., \Y{6,6,6,4,4,1,1}; by enumerating the heights of columns $\Y{h_1,h_2,...,h_{s_1}}$, e.g., \Y{7,5,5,5,3,3}; combining the rows of equal lengths into blocks $\Y{(s_1,p_1),...,(s_N,p_N}$, $p_i$ being the number of rows of the length $s_i$, e.g., \Y{(6,3),(4,2),(1,2)}. \end{tabular}
Each Young diagram defines\footnote{We do not go into details of this correspondence, for systematic presentation see \cite{Barut}.} an irreducible under permutations of indices type of a tensor. There are two main bases that are used for tensors, namely, symmetric and antisymmetric. A tensor $T^{a(s_1),b(s_2),...,d(s_n)}$ is said to have the symmetry of $\Y{s_1,s_2,...,s_n}$, being taken in symmetric basis, iff it is symmetric in each group of indices $a_1...a_{s_1}$, $b_1...b_{s_2}$,..., separately, and the symmetrization of all indices from any group with one index from the next group vanishes, i.e., $T^{a(s_1),...,b(s_i),bc(s_{i+1}-1),...,d(s_n)}\equiv0$. In antisymmetric basis $T^{a[h_1],b[h_2],...,d[h_{s_1}]}$ is antisymmetric in each group of indices $a_1...a_{h_1}$, $b_1...b_{h_2}$, ..., separately, and the anti-symmetrization of all indices from any group with one index from the next group vanishes. To make tensors of the Lorentz algebra irreducible in addition to the Young symmetry conditions a traceless condition has to be imposed, i.e., the contraction of the \lorentz-invariant metric $\eta_{ab}$ with any pair of indices must vanish. It is also required $h_1+h_2\leq d$ since the Young symmetry together with the tracelessness makes \lorentz-tensors with $h_1+h_2>d$ be identically zero. We do not consider (anti)-self dual representations of \lorentz.

\providecommand{\href}[2]{#2}\begingroup\raggedright\endgroup


\begin{thebibliography}{10}

\bibitem{Curtright:1980yk}
T.~Curtright, ``Generalized gauge fields,'' {\em Phys. Lett.} {\bf B165} (1985)
304.

\bibitem{Aulakh:1986cb}
C.~S. Aulakh, I.~G. Koh, and S.~Ouvry, ``Higher spin fields with mixed
  symmetry,'' {\em Phys. Lett.} {\bf B173} (1986)
284.

\bibitem{Ouvry:1986dv}
S.~Ouvry and J.~Stern, ``Gauge fields of any spin and symmetry,'' {\em Phys.
  Lett.} {\bf B177} (1986)
335.

\bibitem{Labastida:1986gy}
J.~M.~F. Labastida and T.~R. Morris, ``Massless mixed symmetry bosonic free
  fields,'' {\em Phys. Lett.} {\bf B180} (1986)
101.

\bibitem{Labastida:1986ft}
J.~M.~F. Labastida, ``Massless bosonic free fields,'' {\em Phys. Rev. Lett.}
  {\bf 58} (1987)
531.

\bibitem{Labastida:1987kw}
J.~M.~F. Labastida, ``Massless particles in arbitrary representation of the
  lorentz group,'' {\em Nucl. Phys.} {\bf B322} (1989)
185.

\bibitem{Metsaev:1995re}
R.~R. Metsaev, ``{Massless mixed symmetry bosonic free fields in d- dimensional
  anti-de Sitter space-time},'' {\em Phys. Lett.} {\bf B354} (1995)
78--84.

\bibitem{Fotopoulos:2008ka}
A.~Fotopoulos and M.~Tsulaia, ``{Gauge Invariant Lagrangians for Free and
  Interacting Higher Spin Fields. A Review of the BRST formulation},''
\href{http://arXiv.org/abs/0805.1346}{{\tt 0805.1346}}.

\bibitem{Metsaev:1993mj}
R.~R. Metsaev, ``{Cubic interaction vertices of totally symmetric and mixed
  symmetry massless representations of the Poincare group in D = 6
  space-time},'' {\em Phys. Lett.} {\bf B309} (1993)
39--44.

\bibitem{Metsaev:1993ap}
R.~R. Metsaev, ``{Generating function for cubic interaction vertices of higher
  spin fields in any dimension},'' {\em Mod. Phys. Lett.} {\bf A8} (1993)
2413--2426.

\bibitem{Metsaev:2005ar}
R.~R. Metsaev, ``{Cubic interaction vertices for massive and massless higher
  spin fields},'' {\em Nucl. Phys.} {\bf B759} (2006) 147--201,
\href{http://arXiv.org/abs/hep-th/0512342}{{\tt hep-th/0512342}}.

\bibitem{Boulanger:2004rx}
N.~Boulanger and S.~Cnockaert, ``{Consistent deformations of (p,p)-type gauge
  field theories},'' {\em JHEP} {\bf 03} (2004) 031,
\href{http://arXiv.org/abs/hep-th/0402180}{{\tt hep-th/0402180}}.

\bibitem{Bekaert:2004dz}
X.~Bekaert, N.~Boulanger, and S.~Cnockaert, ``{No self-interaction for
  two-column massless fields},'' {\em J. Math. Phys.} {\bf 46} (2005) 012303,
\href{http://arXiv.org/abs/hep-th/0407102}{{\tt hep-th/0407102}}.

\bibitem{Bekaert:2006ix}
X.~Bekaert and N.~Boulanger, ``Tensor gauge fields in arbitrary representations
  of gl(d,r). ii: Quadratic actions,'' {\em Commun. Math. Phys.} {\bf 271}
  (2007) 723--773,
\href{http://arXiv.org/abs/hep-th/0606198}{{\tt hep-th/0606198}}.

\bibitem{Francia:2002aa}
D.~Francia and A.~Sagnotti, ``Free geometric equations for higher spins,'' {\em
  Phys. Lett.} {\bf B543} (2002) 303--310,
\href{http://arXiv.org/abs/hep-th/0207002}{{\tt hep-th/0207002}}.

\bibitem{Francia:2002pt}
D.~Francia and A.~Sagnotti, ``{On the geometry of higher-spin gauge fields},''
  {\em Class. Quant. Grav.} {\bf 20} (2003) S473--S486,
\href{http://arXiv.org/abs/hep-th/0212185}{{\tt hep-th/0212185}}.

\bibitem{Bekaert:2002dt}
X.~Bekaert and N.~Boulanger, ``{Tensor gauge fields in arbitrary
  representations of GL(D,R): Duality and Poincare lemma},'' {\em Commun. Math.
  Phys.} {\bf 245} (2004) 27--67,
\href{http://arXiv.org/abs/hep-th/0208058}{{\tt hep-th/0208058}}.

\bibitem{Bekaert:2003az}
X.~Bekaert and N.~Boulanger, ``{On geometric equations and duality for free
  higher spins},'' {\em Phys. Lett.} {\bf B561} (2003) 183--190,
\href{http://arXiv.org/abs/hep-th/0301243}{{\tt hep-th/0301243}}.

\bibitem{Francia:2005bu}
D.~Francia and A.~Sagnotti, ``{Minimal local Lagrangians for higher-spin
  geometry},'' {\em Phys. Lett.} {\bf B624} (2005) 93--104,
\href{http://arXiv.org/abs/hep-th/0507144}{{\tt hep-th/0507144}}.

\bibitem{Bonelli:2003kh}
G.~Bonelli, ``{On the tensionless limit of bosonic strings, infinite symmetries
  and higher spins},'' {\em Nucl. Phys.} {\bf B669} (2003) 159--172,
\href{http://arXiv.org/abs/hep-th/0305155}{{\tt hep-th/0305155}}.

\bibitem{Sagnotti:2003qa}
A.~Sagnotti and M.~Tsulaia, ``{On higher spins and the tensionless limit of
  string theory},'' {\em Nucl. Phys.} {\bf B682} (2004) 83--116,
\href{http://arXiv.org/abs/hep-th/0311257}{{\tt hep-th/0311257}}.

\bibitem{Francia:2006hp}
D.~Francia and A.~Sagnotti, ``{Higher-spin geometry and string theory},'' {\em
  J. Phys. Conf. Ser.} {\bf 33} (2006) 57,
\href{http://arXiv.org/abs/hep-th/0601199}{{\tt hep-th/0601199}}.

\bibitem{Burdik:2001hj}
C.~Burdik, A.~Pashnev, and M.~Tsulaia, ``On the mixed symmetry irreducible
  representations of the poincare group in the brst approach,'' {\em Mod. Phys.
  Lett.} {\bf A16} (2001) 731--746,
\href{http://arXiv.org/abs/hep-th/0101201}{{\tt hep-th/0101201}}.

\bibitem{Burdik:2000kj}
C.~Burdik, A.~Pashnev, and M.~Tsulaia, ``The lagrangian description of
  representations of the poincare group,'' {\em Nucl. Phys. Proc. Suppl.} {\bf
  102} (2001) 285--292,
\href{http://arXiv.org/abs/hep-th/0103143}{{\tt hep-th/0103143}}.

\bibitem{Buchbinder:2007ix}
I.~L. Buchbinder, V.~A. Krykhtin, and H.~Takata, ``Gauge invariant lagrangian
  construction for massive bosonic mixed symmetry higher spin fields,'' {\em
  Phys. Lett.} {\bf B656} (2007) 253--264,
\href{http://arXiv.org/abs/arXiv:0707.2181 [hep-th]}{{\tt arXiv:0707.2181
  [hep-th]}}.

\bibitem{Moshin:2007jt}
P.~Y. Moshin and A.~A. Reshetnyak, ``Brst approach to lagrangian formulation
  for mixed-symmetry fermionic higher-spin fields,'' {\em JHEP} {\bf 10} (2007)
  040,
\href{http://arXiv.org/abs/arXiv:0707.0386 [hep-th]}{{\tt arXiv:0707.0386
  [hep-th]}}.

\bibitem{Skvortsov:2007kz}
E.~D. Skvortsov and M.~A. Vasiliev, ``Transverse invariant higher spin
  fields,''
\href{http://arXiv.org/abs/hep-th/0701278}{{\tt hep-th/0701278}}.

\bibitem{Fronsdal:1978rb}
C.~Fronsdal, ``Massless fields with integer spin,'' {\em Phys. Rev.} {\bf D18}
  (1978)
3624.

\bibitem{Curtright:1979uz}
T.~Curtright, ``Massless field supermultiplets with arbitrary spin,'' {\em
  Phys. Lett.} {\bf B85} (1979)
219.

\bibitem{Vasiliev:1980as}
M.~A. Vasiliev, ``'gauge' form of description of massless fields with arbitrary
  spin,'' {\em Sov. J. Nucl. Phys.} {\bf 32} (1980)
439.

\bibitem{Sorokin:2008tf}
D.~P. Sorokin and M.~A. Vasiliev, ``{Reducible higher-spin multiplets in flat
  and AdS spaces and their geometric frame-like formulation},''
\href{http://arXiv.org/abs/0807.0206}{{\tt 0807.0206}}.

\bibitem{Bengtsson:1986ys}
A.~K.~H. Bengtsson, ``A unified action for higher spin gauge bosons from
  covariant string theory,'' {\em Phys. Lett.} {\bf B182} (1986)
321.

\bibitem{Pashnev:1989gm}
A.~I. Pashnev, ``Composite systems and field theory for a free regge
  trajectory,'' {\em Theor. Math. Phys.} {\bf 78} (1989)
272--277.

\bibitem{Lopatin:1987hz}
V.~E. Lopatin and M.~A. Vasiliev, ``Free massless bosonic fields of arbitrary
  spin in d- dimensional de sitter space,'' {\em Mod. Phys. Lett.} {\bf A3}
  (1988)
257.

\bibitem{Zinoviev:2003ix}
Y.~M. Zinoviev, ``First order formalism for mixed symmetry tensor fields,''
\href{http://arXiv.org/abs/hep-th/0304067}{{\tt hep-th/0304067}}.

\bibitem{Zinoviev:2003dd}
Y.~M. Zinoviev, ``First order formalism for massive mixed symmetry tensor
  fields in minkowski and (a)ds spaces,''
\href{http://arXiv.org/abs/hep-th/0306292}{{\tt hep-th/0306292}}.

\bibitem{Vasiliev:2001wa}
M.~A. Vasiliev, ``Cubic interactions of bosonic higher spin gauge fields in
  ads(5),'' {\em Nucl. Phys.} {\bf B616} (2001) 106--162,
\href{http://arXiv.org/abs/hep-th/0106200}{{\tt hep-th/0106200}}.

\bibitem{Alkalaev:2003qv}
K.~B. Alkalaev, O.~V. Shaynkman, and M.~A. Vasiliev, ``On the frame-like
  formulation of mixed-symmetry massless fields in (a)ds(d),'' {\em Nucl.
  Phys.} {\bf B692} (2004) 363--393,
\href{http://arXiv.org/abs/hep-th/0311164}{{\tt hep-th/0311164}}.

\bibitem{Alkalaev:2005kw}
K.~B. Alkalaev, O.~V. Shaynkman, and M.~A. Vasiliev, ``Lagrangian formulation
  for free mixed-symmetry bosonic gauge fields in (a)ds(d),'' {\em JHEP} {\bf
  08} (2005) 069,
\href{http://arXiv.org/abs/hep-th/0501108}{{\tt hep-th/0501108}}.

\bibitem{Alkalaev:2006rw}
K.~B. Alkalaev, O.~V. Shaynkman, and M.~A. Vasiliev, ``Frame-like formulation
  for free mixed-symmetry bosonic massless higher-spin fields in ads(d),''
\href{http://arXiv.org/abs/hep-th/0601225}{{\tt hep-th/0601225}}.

\bibitem{Skvortsov:2006at}
E.~D. Skvortsov and M.~A. Vasiliev, ``Geometric formulation for partially
  massless fields,'' {\em Nucl. Phys.} {\bf B756} (2006) 117--147,
\href{http://arXiv.org/abs/hep-th/0601095}{{\tt hep-th/0601095}}.

\bibitem{Vasiliev:1988xc}
M.~A. Vasiliev, ``Equations of motion of interacting massless fields of all
  spins as a free differential algebra,'' {\em Phys. Lett.} {\bf B209} (1988)
491--497.

\bibitem{Vasiliev:1988sa}
M.~A. Vasiliev, ``Consistent equations for interacting massless fields of all
  spins in the first order in curvatures,'' {\em Annals Phys.} {\bf 190} (1989)
59--106.

\bibitem{Vasiliev:1992gr}
M.~A. Vasiliev, ``Unfolded representation for relativistic equations in (2+1)
  anti-de sitter space,'' {\em Class. Quant. Grav.} {\bf 11} (1994)
649--664.

\bibitem{Sullivan77}
D.~Sullivan, ``Infinitesimal computations in topology,'' {\em Publ. Math. IHES}
  {\bf 47} (1977) 269--331.

\bibitem{D'Auria:1982nx}
R.~D'Auria and P.~Fre, ``Geometric supergravity in d = 11 and its hidden
  supergroup,'' {\em Nucl. Phys.} {\bf B201} (1982)
101--140.

\bibitem{D'Auria:1982pm}
R.~D'Auria, P.~Fre, P.~K. Townsend, and P.~van Nieuwenhuizen, ``Invariance of
  actions, rheonomy and the new minimal n=1 supergravity in the group manifold
  approach,'' {\em Ann. Phys.} {\bf 155} (1984)
423.

\bibitem{Nieuwenhuizen:1982zf}
P.~van Nieuwenhuizen, ``Free graded differential superalgebras,''. Invited talk
  given at 11th Int. Colloq. on Group Theoretical Methods in Physics, Istanbul,
  Turkey, Aug 23- 28, 1982.

\bibitem{Fre:2005px}
P.~Fre, ``{M-theory FDA, twisted tori and Chevalley cohomology},'' {\em Nucl.
  Phys.} {\bf B742} (2006) 86--123,
\href{http://arXiv.org/abs/hep-th/0510068}{{\tt hep-th/0510068}}.

\bibitem{Fre:2008qw}
P.~Fre and P.~A. Grassi, ``{Free Differential Algebras, Rheonomy, and Pure
  Spinors},''
\href{http://arXiv.org/abs/0801.3076}{{\tt 0801.3076}}.

\bibitem{Vasiliev:1989yr}
M.~A. Vasiliev, ``Dynamics of massless higher spins in the second order in
  curvatures,'' {\em Phys. Lett.} {\bf B238} (1990)
305--314.

\bibitem{Vasiliev:1990en}
M.~A. Vasiliev, ``Consistent equation for interacting gauge fields of all spins
  in (3+1)-dimensions,'' {\em Phys. Lett.} {\bf B243} (1990)
378--382.

\bibitem{Vasiliev:2003ev}
M.~A. Vasiliev, ``Nonlinear equations for symmetric massless higher spin fields
  in (a)ds(d),'' {\em Phys. Lett.} {\bf B567} (2003) 139--151,
\href{http://arXiv.org/abs/hep-th/0304049}{{\tt hep-th/0304049}}.

\bibitem{Konshtein:1988yg}
S.~E. Konstein and M.~A. Vasiliev, ``Massless representations and admissibility
  condition for higher spin superalgebras,'' {\em Nucl. Phys.} {\bf B312}
  (1989)
402.

\bibitem{Skvortsov:2008vs}
E.~D. Skvortsov, ``{Mixed-Symmetry Massless Fields in Minkowski space
  Unfolded},''
\href{http://arXiv.org/abs/0801.2268}{{\tt 0801.2268}}.

\bibitem{Vasiliev:2007yc}
M.~A. Vasiliev, ``On conformal, sl(4,r) and sp(8,r) symmetries of 4d massless
  fields,''
\href{http://arXiv.org/abs/arXiv:0707.1085 [hep-th]}{{\tt arXiv:0707.1085
  [hep-th]}}.

\bibitem{Bekaert:2005vh}
X.~Bekaert, S.~Cnockaert, C.~Iazeolla, and M.~A. Vasiliev, ``Nonlinear higher
  spin theories in various dimensions,''
\href{http://arXiv.org/abs/hep-th/0503128}{{\tt hep-th/0503128}}.

\bibitem{Vasiliev:1986td}
M.~A. Vasiliev, ``Free massless fields of arbitrary spin in the de sitter space
  and initial data for a higher spin superalgebra,'' {\em Fortsch. Phys.} {\bf
  35} (1987)
741--770.

\bibitem{Shaynkman:2000ts}
O.~V. Shaynkman and M.~A. Vasiliev, ``Scalar field in any dimension from the
  higher spin gauge theory perspective,'' {\em Theor. Math. Phys.} {\bf 123}
  (2000) 683--700,
\href{http://arXiv.org/abs/hep-th/0003123}{{\tt hep-th/0003123}}.

\bibitem{Barut}
A.~O. Barut and R.~Raczka, ``Theory of group representations and
  applications,''. Singapore, Singapore: World Scientific ( 1986) 717p.

\end{thebibliography}
\end{document}